\documentclass[11pt]{article}

\usepackage{latexsym,color,graphicx,amsmath,amssymb}

\usepackage{enumitem}

\def\scirc#1{${\small$\circ$}$}

\def\R{{\mathbb R}}
\def\reals{{\mathbb R}}

\def\C{{\mathbb C}}
\def\Z{{\mathbb Z}}
 
\newtheorem{thm}{Theorem}%[Section]
\newtheorem{lem}[thm]{Lemma}
\newtheorem{prop}[thm]{Proposition}
\newtheorem{cor}[thm]{Corollary}
\newtheorem{clm}[thm]{Claim}
 
\begin{document}

\title{Polynomials vanishing on grids: The Elekes-R\'onyai problem revisited\thanks{%
Work on this paper by Orit E. Raz and Micha Sharir was supported by Grant 892/13 from the Israel Science Foundation. Work by Micha Sharir was also supported by Grant 2012/229 from the U.S.--Israel Binational Science Foundation, by the Israeli Centers of Research Excellence (I-CORE) program (Center No.~4/11), and by the Hermann Minkowski-MINERVA Center for Geometry at Tel Aviv University.  Work by J\'ozsef Solymosi was supported by NSERC, ERC-AdG 321104, and OTKA NK 104183 grants. }}
%\date{}
\author{
Orit E. Raz\thanks{%
School of Computer Science, Tel Aviv University,
Tel Aviv 69978, Israel.
{\sl oritraz@post.tau.ac.il} }
\and
Micha Sharir\thanks{%
School of Computer Science, Tel Aviv University,
Tel Aviv 69978, Israel.
{\sl michas@post.tau.ac.il} }
\and
J\'ozsef Solymosi\thanks{%
Department of Mathematics,
University of British Columbia,
Vancouver, BC, V6T 1Z4, Canada. 
\newline {\sl solymosi@math.ubc.ca}} }

\maketitle

\begin{abstract}
In this paper we characterize real bivariate polynomials which have a small range over large Cartesian products. We show that for every constant-degree bivariate real polynomial $f$, either $|f(A,B)|=\Omega(n^{4/3})$, for every pair of finite sets $A,B\subset\R$, with $|A|=|B|=n$ (where the constant of proportionality depends on $\deg f$), or else $f$ must be of one of the special forms $f(u,v)=h(\varphi(u)+\psi(v))$, or $f(u,v)=h(\varphi(u)\cdot\psi(v))$, for some univariate polynomials $\varphi,\psi,h$ over $\R$. This significantly improves a result of Elekes and R\'onyai~\cite{ER00}.

Our results are cast in a more general form, in which we give an upper bound for the number of zeros of $z=f(x,y)$ on a triple Cartesian product $A\times B\times C$, when the sizes $|A|$, $|B|$, $|C|$ need not be the same; the upper bound is $O(n^{11/6})$ when $|A|=|B|=|C|=n$, where the constant of proportionality depends on $\deg f$, unless $f$ has one of the aforementioned special forms.

This result provides a unified tool for improving bounds in various Erd\H os-type problems in geometry and additive combinatorics. Several applications of our results to problems of these kinds are presented. For example, we show that the number of distinct distances between $n$ points lying on a constant-degree parametric algebraic curve which does not contain a line, in any dimension, is $\Omega(n^{4/3})$, extending the result of Pach and de Zeeuw~\cite{PdZ} and improving the bound of Charalambides~\cite{Cha}, for the special case where the curve under consideration has a polynomial parameterization. We also derive improved lower bounds for several variants of the sum-product problem in additive combinatorics.
\end{abstract}


%-------------------------------Section:Introduction---------------------------

\section{Introduction}

%-----------------------------Background----------------------------------------
\subsection{Background}
In 2000, Elekes and R\'onyai~\cite{ER00} considered the following problem. Let $A,B$ be two sets, each of $n$ real numbers, and let $f$ be a real bivariate polynomial of some constant degree. They showed that if $|f(A\times B)|\le cn$, for some constant $c$ that depends on $\deg f$, and for $n\ge n_0(c)$, for sufficiently large threshold $n_0(c)$ that depends on $c$, then $f$ must be of one of the special forms $f(u,v)=h(\varphi(u)+\psi(v))$, or $f(u,v)=h(\varphi(u)\cdot\psi(v))$, for some univariate polynomials $\varphi,\psi,h$ over $\R$.

In a variant of this setup, we are given, in addition to $A, B$ and $f$, another set $C$ of $n$ real numbers, and the quantity $|f(A\times B)|$ is replaced by the number $M$ of triples $(a,b,c)\in A\times B\times C$ such that $c=f(a,b)$. Elekes and R\'onyai have shown that if $M=\Omega(n^2)$ then $f$ must have one of the above special forms. Elekes and Szab\'o~\cite{ESz} were able to extend this theorem to implicit surfaces $F(x, y, z) = 0$, and also showed that, unless $F$ has a certain specific special form (see \cite{ESz} for precise formulation and more details), the surface can only contain $O(n^{2-\eta})$ points of $A\times B\times C$, for some exponential `gap' $\eta > 0$ that depends on the degree of the polynomial $F$ (they do not make the values of $\eta$ explicit, and point out that it is `rather small'). The study of Elekes and Szab\'o also considers more involved setups, where $A$, $B$, $C$, and $F$ are embedded in higher dimensions, and/or the underlying field is the complex field $\C$. 

In this paper, we prove that, unless $f$ has one of the aforementioned special forms, $M=O(n^{11/6})$ (where the constant of proportionality depends on $\deg f$). In doing so, we give two alternative proofs of this result, which we believe to be simpler than the ones in~\cite{ER00,ESz}. Our result improves the previous ones, by making the bound on $M$ explicit, with an exponent that is independent of the degree of $f$. (The previous gap is only given in \cite{ESz}; the former paper \cite{ER00} only shows that $M=o(n^2)$.)

We actually establish a more general result than that of \cite{ER00}, for the case where $|A|,|B|,|C|$ are not necessarily equal. The threshold bound $O(n^{11/6})$ is then replaced by a more involved expression in $|A|,|B|,|C|$ (see Theorem~\ref{main2} below). This generalization requires a more careful and somewhat more involved analysis. Schwartz, Solymosi and de Zeeuw~\cite{SSdZ} have recently considered the special `unbalanced' case where $|A|=|C|=n$ and $|B|=n^{1/2+\varepsilon}$, for any fixed $\varepsilon>0$, and showed that the graph of $f$ must contain $o(n^{3/2+\varepsilon})$ points of $A\times B\times C$, unless $f$ is of one of the special forms. Our analysis applies in this setup, and slightly improves (and makes more concrete) the bound just mentioned.

The technique used in this paper has some common features with the one used in \cite{ER00}. A discussion of the similarities and differences between the two approaches is given in the concluding section (Section~\ref{sec:con}).

Besides being an interesting problem in itself, the Elekes-R\'onyai setup, and certain generalizations thereof, such as those considered by Elekes and Szab\'o~\cite{ESz}, arise in many problems in combinatorial geometry. This connection has resurfaced in several recent works, including problems on distinct distances in several special configurations (see Sharir et al.~\cite{SSS} and Pach and de Zeeuw~\cite{PdZ} for ad-hoc treatments of these instances). In many of these problems it is essential to allow the sets $A$, $B$, and $C$ under consideration to be of different sizes (usually, the interest is then in estimating the cardinality of one of these sets), and then it useful to have the unbalanced version of% Theorem \ref{main2}. To demonstrate this (beyond the refinement of the aforementioned result in \cite{SSdZ}), let us describe one of these problems in more detail; other applications will be given in Section~\ref{sec:appl}.

\paragraph{Distinct distances between two lines.} 

Consider the following special instance of the distinct distances problem of Erd\H{o}s. Let $\ell_1,\ell_2$ be two lines in the plane which are neither parallel nor orthogonal, and let $P_i$ be a finite set of points on $\ell_i$, for $i=1,2$. Sharir et al.~\cite{SSS} have recently shown that the number of distinct distances between pairs in $P_1\times P_2$ is
$$
\Omega\left( \min\left\{ |P_1|^{2/3}|P_2|^{2/3},\;|P_1|^2,\;|P_2|^2 \right\}\right).
$$

To see the connection with the Elekes-R\'onyai setup, let $D$ denote the set of all squared distinct distances determined by $P_1\times P_2$, and consider the function $F:\ell_1\times\ell_2\to\R$, given by $F(p,q)=\|p-q\|^2$. Let $M$ denote the number of triples $(p,q,d)\in P_1\times P_2\times D$, for which $d=F(p,q)$. By the definition of $D$, we have $M=|P_1||P_2|$. Thus an upper bound on $M$ (in terms of $|P_1|,|P_2|,$ and $|D|$) would yield a lower bound on $|D|$. This is essentially the setup in Theorem \ref{main2}, if we regard $\ell_1$ and $\ell_2$ as two copies of $\R$, so that $F$ becomes a quadratic bivariate polynomial over $\R$. Then, by Theorem \ref{main2}, stated below, either $f$ is of one of the special forms specified in the theorem, which can be shown not to be the case, or else
$$
|P_1||P_2|=M=O\left( \left(|P_1|^{2/3}|P_2|^{2/3}+|P_1|+|P_2|\right)|D|^{1/2}\right),
$$
which implies that
\begin{equation}\label{eq:dist}
|D|=\Omega\left( \min\left\{ |P_1|^{2/3}|P_2|^{2/3},\;|P_1|^2,\;|P_2|^2 \right\}\right),
\end{equation}
which is exactly the bound obtained in \cite{SSS}. (There are several simple ways to show that $f$ is not of one of the specific forms, which we omit in this quick discussion.)

\paragraph{Extensions.} The high-level approach used in this paper can be viewed as an instance of a more general technique, applicable to geometric problems that involve an interaction between three sets of real numbers, where the interaction can be expressed by a general trivariate (constant-degree) polynomial equation $F(x,y,z)=0$. This is very much related to the setup considered by Elekes and Szab\'o~\cite{ESz}, and we will discuss the issues involved in this extension at the end of the paper. Such an extension would facilitate further applications of the new machinery, to a variety of problems of this kind. 

Two recent studies, by Solymosi and Sharir~\cite{SS} and by Raz et al.~\cite{RSS}, involve problems of this form. The former paper studies the problem of obtaining a lower bound on the number of distinct distances between three non-collinear points and $n$ other points in the plane (the lower bound obtained there is $\Omega(n^{6/11})$). The latter paper reconsiders the problem, previously studied by Elekes et al.~\cite{ESSz}, of obtaining an upper bound on the number of triple intersection points between three families of $n$ unit circles, where all the circles of the same family pass through a fixed point in the plane (the upper bound obtained there is $O(n^{11/6})$). In both cases the analysis follows a general paradigm, similar to the one in this paper (except that the underlying polynomial is trivariate rather than bivariate), and faces a technical issue that is handled by problem-specific  ad-hoc techniques. This issue, which we do not yet spell out, will become clear after digesting our analysis, and will be discussed in the concluding section.

%--------------------------------------Our results---------------------------------
\subsection{Our results}
Our main result, for the case $|A|=|B|=|C|$, is as follows.
\begin{thm}\label{main}
Let $A$, $B$, and $C$ be three finite sets of real numbers, each of cardinality $n$. Let $f\in\R[u,v]$ be a bivariate real polynomial of constant degree $d\ge 2$, and let $M$ denote the number of intersection points of the surface $w=f(u,v)$ with $A\times B\times C$ in $\R^3$. Then either $M= O\left(n^{11/6}\right)$, where the constant of proportionality depends on $\deg f$, or $f$ is of one of the forms $f(u,v)=h(\varphi(u)+\psi(v))$, or $f(u,v)=h(\varphi(u)\cdot\psi(v))$, for some univariate polynomials $\varphi,\psi,h$ over $\R$.
\end{thm}

As mentioned earlier, in some applications the sets $A,B,C$ are not of the same cardinality. As promised, our analysis caters to these asymmetric situations too, and establishes the following more general result.   
\begin{thm}\label{main2}
Let $A$, $B$, and $C$ be three finite sets of real numbers. Let $f\in\R[u,v]$ be a bivariate real polynomial of constant degree $d\ge 2$, and let $M$ denote the number of intersection points of the surface $w=f(u,v)$ with $A\times B\times C$ in $\R^3$. Then either 
\begin{align*}
M= O\left(\min\left\{|A|^{2/3}|B|^{1/2}|C|^{2/3}+|A|^{1/2}|B|^{2/3}|C|^{2/3}+|A|+|B|,\right.\right.\\
\left.\left.\left(|A|^{2/3}|B|^{2/3}+|A|+|B|\right)|C|^{1/2}\right\}\right),
\end{align*}
again where the constant of proportionality depends on $\deg f$, or $f$ is of one of the forms $f(u,v)=h(\varphi(u)+\psi(v))$, or $f(u,v)=h(\varphi(u)\cdot\psi(v))$, for some univariate polynomials $\varphi,\psi,h$ over $\R$.
\end{thm}

The following is an immediate consequence of the second part of the bound in Theorem~\ref{main2}, which suffices for many of our applications. It is obtained by putting $C=f(A,B)$ and $M=|A||B|$. We do expect, though, that further applications will need to exploit the full generality of our bounds.
\begin{cor}\label{cor:main}
Let $A,B\subset\R$ be two finite sets, and let $f$ be a bivariate constant-degree real polynomial. Then, unless $f$ is of one of the special forms specified in the statement of Theorem \ref{main2}, we have
$$
|f(A,B)|= \Omega\left( \min\left\{ |A|^{2/3}|B|^{2/3},\;|A|^2,\;|B|^2 \right\}\right).
$$
\end{cor}

We prove only Theorem \ref{main2}, and do it in two parts, respectively establishing the first expression (in Section~\ref{sec:part1}) and the second expression (in Section~\ref{sec:part2}) in the asserted bound. (Either of these proofs in itself suffices to obtain Theorem~\ref{main} in the balanced case $|A|=|B|=|C|$, so there is no need to digest both proofs for this special case, but they provide different bounds and cater to different ranges of $|A|$, $|B|$, and $|C|$ in the unbalanced case.)

In Section~\ref{sec:appl} we present several applications of our result. Most of these problems have already been considered in the literature, but our machinery yields improved bounds, and simplifies some of the earlier proofs. These applications include (i) improved lower bounds on the number of distinct slopes determined by points on a curve, and on the number of distinct distances determined by such points, and (ii) improved lower bounds for variants of the sum-product problem.

%----------------------------------------------Preliminaries--------------------------------------------------------------------
\section{Preliminaries}\label{sec:pre}
\paragraph{Algebraic preliminaries.}
Let $K$ be a field, and let $p$ be a bivariate polynomial with coefficients in $K$. We say that $p$ is \emph{decomposable} over $K$ if we can write $p(u,v)=r${\small$\circ$}$q(u,v)=r(q(u,v))$, where $r$ is a univariate polynomial of degree at least two, and $q$ is a bivariate polynomial, both with coefficients in $K$. Otherwise, $p$ is said to be \emph{indecomposable} (over $K$). It is easy to see that a decomposable polynomial $p$ over $K$ is reducible over $\bar{K}$, where $\bar K$ stands for the algebraic closure of $K$. Indeed, if $p=r${\small$\circ$}$q$, where $r$ and $q$ are as before, then $p(u,v)=\prod_i(q(u,v)-z_i)$, where $z_i$, $i=1,\ldots,\deg r$, are the roots of $r$ (which is indeed a non-trivial factorization since $\deg r\ge2$).

The following theorem of Stein~\cite{St} is crucial for our analysis. (See Shen~\cite{She} for another recent application of Stein's theorem to a related problem.) It is concerned with the connection between the decomposability of $p$ and the reducibility of $p-\lambda$, for elements $\lambda\in \bar K$.
\begin{thm}[Stein~\cite{St}] \label{St}
Let $\bar K$ be an algebraically closed field, and let $p$ be a bivariate polynomial with coefficients in $\bar K$. If $p$ is indecomposable over $\bar K$, then
$$
\left|\{\lambda\in \bar K\mid p-\lambda~\text{is reducible over}~\bar K\}\right|<\deg p.
$$
\end{thm}
The requirement in Theorem \ref{St}, that the field under consideration be algebraically closed, is not essential, as shown by the following theorem, taken from Ayad~\cite[Theorems 4 and 7]{Aya}.
\begin{thm}[Ayad~\cite{Aya}]\label{Aya}
Let $K$ be a field of characteristic zero, and let $p$ be a bivariate polynomial with coefficients in $K$. Then $f$ is decomposable over $K$ if and only if it is decomposable over $\bar K$.
\end{thm}
Combining Theorem \ref{St} and Theorem \ref{Aya}, we obtain the following corollary, which is formulated specifically for our needs in the proof of Theorem \ref{main}.
\begin{cor}\label{cordecom}
Let $p$ be a bivariate polynomial in $\R[x,y]$. If $p$ is indecomposable over $\R$, then
$$
\left|\{\lambda\in \R\mid p-\lambda~\text{is reducible over}~\R\}\right|<\deg p.
$$
\end{cor}

We also make use of the classical bivariate B\'ezout's theorem (see, e.g., \cite{CLO}), again specialized to real polynomials.
\begin{thm}[B\'ezout]\label{bezout}
Let $f$ and $g$ be two bivariate polynomials over $\R$, with degrees $d_f$ and $d_g$, respectively. If $f$ and $g$ vanish simultaneously at more than $d_f d_g$ points of $\R^2$, then $f$ and $g$ have a common non-trivial factor.
\end{thm}

The following result is useful in analyzing the zero set of a bivariate polynomial on a grid. It is a specialization to two dimensions of the more general result presented in \cite{Sch80} and \cite{Zi89}.
\begin{lem}[Schwartz-Zippel Lemma~\cite{Sch80,Zi89}]\label{SZ}
Let $g$ be a real bivariate polynomial of degree $\delta$, and let $U,V$ be two finite point sets in $\R^2$, with $|U|=|V|=n$. Then $g$ has at most $\delta n$ zeros in $U\times V$. In case $|U|\neq|V|$, this number is $\min\{\delta|U|,\delta|V|\}$.
\end{lem}

\paragraph{Combinatorial preliminaries.}
One of the main ingredients of the proof of Theorem \ref{main2} will be a reduction to a problem involving incidences between points and curves in the plane. We therefore recall some basic results in incidence theory, which has its roots in the following classical result of Szemer\'edi and Trotter~\cite{ST}.
%-------------thm:Szemeredi-Trotter--------------------
\begin{thm}[Szemer\'edi-Trotter~\cite{ST}]\label{ST}
The number of incidences between $m$ distinct points and $n$ distinct lines in $\R^2$ is $O(m^{2/3}n^{2/3} + m + n)$.
\end{thm}
Theorem \ref{ST} has seen a number of generalizations. For example, we have:
%-----------------thm:Pach-Sharir-----------------
\begin{thm}[Pach-Sharir~\cite{PS98}]\label{PS}
Let $P$ be a collection of $m$ distinct points in $\R^2$ and $S$ a collection of $n$ distinct curves with $k$ degrees of freedom, i.e., there exists a constant $C_0$ such that any two curves can meet in at most $C_0$ points and at most $C_0$ curves can contain any $k$ given points. Then the number of incidences between the points of $P$ and the curves of $S$ is
$O\left(m^{\frac{k}{2k-1}}n^{\frac{2k-2}{2k-1}}+m+n\right)$, 
where the implicit constant depends only on $C_0$ and $k$.
\end{thm}
(In the Szemer\'edi-Trotter setup, $k=2$.) Theorem \ref{PS} was proved using the Crossing Lemma of Ajtai et al.~and of Leighton (see, e.g., \cite{PA95} for a more recent exposition), which provides a lower bound for the edge-crossing number for graphs embedded in the plane.  It was first employed in incidence geometry by Sz\'ekely~\cite{Sz97}, where, among other results, it has yielded a simple and elegant proof of the Szemer\'edi-Trotter theorem.
\begin{thm}[Crossing Lemma]\label{thm:cross}
Let $G=(V,E)$ be a simple graph drawn in the plane. Then
$$
|E|=O\left(|V|+|V|^{2/3}{\rm Cr}(G)^{1/3}\right),
$$
where ${\rm Cr}(G)$ is the number of pairs $(e,e')$ of edges of $E$, such that the drawing of $e$ and $e'$ cross each other.
\end{thm}

%----------------------------------------------------PART ONE-----------------------------------------------------------------------------------------------------------------------
\section{Proof of Theorem \ref{main2}: Part 1}\label{sec:part1}
The proof is given in two installments, each establishing (when $f$ does not have one of the special forms) a different upper bound on $M$; the combination of these bounds yields Theorem \ref{main2}. The first part is presented in this section, and the second part in Section~\ref{sec:part2}. As noted, when $|A|=|B|=|C|$, both parts of the proof yield the same bound, and there is no need to have both.
\begin{prop}\label{prop:part1}
Let $A$, $B$, and $C$ be three finite sets of real numbers. Let $f\in\R[u,v]$ be a bivariate real polynomial of constant degree $d\ge 2$, and let $M$ denote the number of intersection points of the surface $w=f(u,v)$ with $A\times B\times C$ in $\R^3$. Then either 
$$
M= O\left(|A|^{2/3}|B|^{1/2}|C|^{2/3}+|A|^{1/2}|B|^{2/3}|C|^{2/3}+|A|+|B|\right),
$$
with a constant of proportionality that depends on $d$, or $f$ is of one of the forms $f(u,v)=h(\varphi(u)+\psi(v))$, or $f(u,v)=h(\varphi(u)\cdot\psi(v))$, for some real univariate polynomials $\varphi,\psi,h$.
\end{prop}
\noindent{\bf Proof.}
It suffices to consider only values $a\in A$ for which $f(a,v)$ is non-constant (regarded as a polynomial in $v$). Indeed, there are at most $d_u$ values of $a$, for which $f(a,v)$ is independent of $v$, each determines a unique value $c$ (possibly in $C$) such that $f(a,v)\equiv c$. Hence the number of triples $(a,b,c)\in A\times B\times C$ for which $a$ is problematic (in the above sense) and $f(a,b)=c$ is at most $d_u\cdot |B|$, which is subsumed in the asserted bound on $M$. Symmetrically, the number of triples $(a,b,c)\in A\times B\times C$ for which $f(u,b)$ is independent of $u$ and $f(a,b)=c$ is at most $d_v\cdot |A|$, which is again subsumed in the asserted bound on $M$. To recap, by trimming $A$ and $C$ accordingly, we may assume that (i) for each $a\in A$, $f(a,z)$ is non-constant in $z$, and (ii) for each $c\in C$, no value $z_0\in\R$ yields a constant polynomial $f(u,z_0)$ (i.e., independent of $u$) whose value is $c$.

We first consider the case where $f$ is indecomposable. We put $d_u=\deg_u(f)$ and $d_v=\deg_v(f)$ (so $d\le d_u+d_v$). With each pair $(a,c)\in A\times C$, we associate a curve $\bar\gamma_{a,c}$ in $\R^3$, defined as
\begin{equation}\label{barcurve}
\bar\gamma_{a,c}:=\{(x,y,z)\in\R^3\mid y=f(a,z)\land c=f(x,z)\}.
\end{equation}
$\bar\gamma_{a,c}$ is the intersection curve of the two cylindrical surfaces 
$$
\sigma_{a}:=\{(x,y,z)\in\R^3\mid y=f(a,z)\}\quad\text{and}\quad\sigma_{c}^*:=\{(x,y,z)\in\R^3\mid c=f(x,z)\}
$$
in $\R^3$. To see that this is indeed a (one-dimensional) curve, note that for every value of $z$, $y$ is determined uniquely by the equation $y=f(a,z)$, and there are at most $d$ values of $x$ for which $c=f(x,z)$; this follows from the trimming of $C$ used above. Hence the intersection $\bar\gamma_{a,c}$ cannot be two dimensional. 

Note that there are at most $d_u$ pairs $(a,c)\in A\times C$ that are associated with the same curve. Indeed, let $\bar\gamma$ be some curve of the form~(\ref{barcurve}), and let $(x,y,z)$ be a generic point of $\bar\gamma$. A pair $(a,c)$ which is associated with $\bar\gamma$ satisfies $y=f(a,z)$ and $c=f(x,z)$. This clearly determines $c$ uniquely, and $a$ is one of the at most $d_u$ roots of $y=f(a,z)$, regarded as a polynomial in $a$. (Again, our trimming of $A$ guarantees that $f(a,z)$ is not a constant polynomial for any $a\in A$.) 

We let $\gamma_{a,c}$ denote the projection of $\bar\gamma_{a,c}$ onto the $xy$-plane in $\R^3$, which we identify with $\R^2$. In other words, $\gamma_{a,c}$ is the locus of all points $(x,y)\in\R^2$ for which there exists $z\in\R$, such that $y=f(a,z)$ and $c=f(x,z)$. 

Let $\Gamma:=\{\gamma_{a,c}\mid (a,c)\in A\times C\}$ denote the multiset of these curves, allowing for the possibility that the same projection might be shared by more than one original curve, even though the original curves themselves are (up to a constant multiplicity) distinct, as argued above, and let $I$ denote the number of incidences between the curves of $\Gamma$ and the points of $\Pi:=A\times C$; since the curves of $\Gamma$ can potentially overlap or coincide, we count incidences with multiplicity: A point lying on $k$ coinciding curves (or, more precisely, on an irreducible component shared by $k$ of the curves) contributes $k$ to the count $I$.

Recall that $M$, as defined in the theorem, is the number of intersection points of the surface\footnote{Note that the roles of the $y$-axis and the $z$-axis are reversed in the present setup.} $y=f(x,z)$ with the point set $A\times C\times B$ in $\R^3$. We obtain an upper bound on $M$ as follows. For each $b\in B$, put
$$
\Pi_b=\left(A\times C\right)_{b}:=\{(a,c)\in A\times C\mid c=f(a,b)\},
$$
and put $M_{b}=|\Pi_b|=\left|\left(A\times C\right)_{b}\right|$. We clearly have $M=\sum_{b\in B}M_b$.

Fix $b\in B$, and note that for any pair of pairs $(a_1,c_1)$, $(a_2,c_2)\in \Pi_{b}$, we have $(a_1,c_2)\in\gamma_{a_2,c_1}$ and $(a_2,c_1)\in\gamma_{a_1,c_2}$. Moreover, for a fixed pair $(a_1,c_1)$, $(a_2,c_2)$ of this kind, the number of values $b$ for which $(a_1,c_1)$ and $(a_2,c_2)$ both belong to $\Pi_{b}$ is at most the constant $v$-degree $d_v$ of $f$, unless $w-f(u,v)$ vanishes identically on the two lines $(a_1,c_1)\times\R$, $(a_2,c_2)\times\R$. However, the latter situation cannot arise because of our trimming of $A$ and $C$. It then follows, using the Cauchy-Schwarz inequality, that 
\begin{align}\label{cs}
M=\sum_{b\in B} M_{b}&\le\left(\sum_{b\in B} M_b^2\right)^{1/2}\cdot|B|^{1/2}\\
&\le\left( d_vI+d_u^2|B|\right)^{1/2}|B|^{1/2}=O\left(I^{1/2}|B|^{1/2}+|B|\right).\nonumber
\end{align}
Hence deriving an upper bound on $I$ would yield an upper bound on $M$. Bounding $I$ is an instance of a fairly standard point-curve incidence problem, which can in principle be tackled using the well established machinery reviewed in Section \ref{sec:pre}. However, to apply this machinery, it is essential for the curves of $\Gamma$ to have a constant bound on their multiplicity. More precisely, we need to know that no more than $O(1)$ curves of $\Gamma$ can share a common irreducible component. When this is indeed the case, we derive an upper bound on the number of incidences, using the following proposition, whose proof is deferred to Section \ref{subsec:incid}. 
%-----------------------------
\begin{prop}\label{incid}
Let $\Gamma$ and $\Pi$ be as above, and assume that no more than $m_0:=d_ud_v+d_ud(d+d_v-1)$ curves of $\Gamma$ can share an irreducible component. Then the number $I$ of incidences between $\Gamma$ and $\Pi$ is $O(|\Gamma|^{2/3}|\Pi|^{2/3}+|\Gamma|+|\Pi|)$, where the constant of proportionality depends on $d$.
\end{prop}
%----------------------------
Since $|\Pi| = |\Gamma| = |A||C|$, it follows that in this case $I=O\left(|A|^{4/3}|C|^{4/3}\right)$. Plugging this bound into (\ref{cs}), we get $M=O\left(|A|^{2/3}|B|^{1/2}|C|^{2/3}+|B|\right)$.

In the complementary case, namely when there exist $m>m_0$ curves of $\Gamma$ that share an irreducible curve, we start all over again, with the roles of the variables $u,v$ of $f$ switched. Although the analysis is fully symmetric, we spell out a few details in the interest of clarity. We now associate with each pair $(b,c)\in B\times C$ a curve $\bar\gamma_{b,c}$ in $\R^3$, defined as
$$
\bar\gamma_{b,c}:=\{(x,y,z)\in\R^3\mid y=f(z,b)\land c=f(z,x)\}.
$$
We let $\gamma_{b,c}$ denote the projection of $\bar\gamma_{b,c}$ onto the $xy$-plane in $\R^3$, which we identify, as above, with $\R^2$. Then $\gamma_{b,c}$ is the locus of all points $(x,y)\in\R^2$ for which there exists $z\in\R$, such that $y=f(z,b)$ and $c=f(z,x)$. We let $\tilde\Gamma:=\{\gamma_{b,c}\mid (b,c)\in B\times C\}$ denote the multiset of the projected curves, and let $\tilde I$ denote the number of incidences (again, counted with multiplicity) between the curves of $\tilde\Gamma$ and the points of $\tilde\Pi:=B\times C$.

With this shuffling of coordinates, $M$ is now the number of intersection points of the surface $y=f(z,x)$ with the point set $B\times C\times A$ in $\R^3$. If no more than $\tilde m_0:=d_ud_v+d_vd(d+d_u-1)$ curves of $\tilde\Gamma$ can share a common irreducible component (note that the roles of $d_u$ and $d_v$ are switched, as they should be, in the definition of $\tilde m_0$), we apply Proposition \ref{incid} to $\tilde\Gamma$ and $\tilde \Pi$ and derive an upper bound on $\tilde I$. The analysis is fully symmetric to the one given above, and yields 
$$
M=O\left(\tilde I^{1/2}|A|^{1/2}+|A|\right)=O\left(|A|^{1/2}|B|^{2/3}|C|^{2/3}+|A|\right).
$$

Thus we have proved the following lemma.
%--------Lemma
\begin{lem}~\label{lem1}
If either (i) no irreducible curve is a component of more than $m_0$ curves of $\Gamma$, or (ii) no irreducible curve is a component of more than $\tilde m_0$ curves of $\tilde\Gamma$, with $m_0,\tilde m_0$ as above, we have
$$
M=O\left(|A|^{2/3}|B|^{1/2}|C|^{2/3}+|A|^{1/2}|B|^{2/3}|C|^{2/3}+|A|+|B|\right).
$$
\end{lem}
%--------
Note that this is the bound asserted in Proposition~\ref{prop:part1}. It thus remains to consider the case where both $\Gamma$ and $\tilde\Gamma$ contain curves of large multiplicity, in the precise sense formulated in Lemma~\ref{lem1}. We show that in this case $f$ must have one of the special forms asserted in the theorem. (More precisely, since we are still under the assumption that $f$ is indecomposable, the analysis yields a more restricted representation of $f$; see below for more details.) The following proposition ``almost" brings us to those forms.
%-------------------Proposition
\begin{prop}\label{weakstruct}
Suppose that there exists an irreducible algebraic curve in $\R^2$ that is shared by more than $m_0=d_ud_v+d_ud(d+d_v-1)$ distinct curves of $\Gamma$. Then $f$ is of the form 
\begin{equation}\label{weakxy}
f(u,v)=up(u)q(v)+r(v),
\end{equation}
for some real univariate polynomials $p,q,r$.
\end{prop}
%---------------------
The proof of Proposition \ref{weakstruct} is given in Section \ref{subsec:weakstruct}; it will exploit our (temporary) assumption that $f$ is indecomposable. Applying a symmetric version of Proposition \ref{weakstruct}, in which the roles of $A$ and $B$, and the respective $x$- and $z$-coordinates, are switched, we conclude that we also have
\begin{equation}\label{weakyx}
f(u,v)=v\tilde p(v)\tilde q(u)+\tilde r(u),
\end{equation}
for suitable real univariate polynomials $\tilde p, \tilde q,\tilde r$.

Equating the two expressions (\ref{weakxy}) and (\ref{weakyx}), and substituting $u=0$ (resp., $v=0$), we get
\begin{align*}
r(v) & = v\tilde p(v)\tilde q(0) + \tilde r(0) \\
\tilde r(u) &= up(u)q(0)+r(0).
\end{align*}
That is,
$$
up(u)q(v)+v\tilde p(v)\tilde q(0)+\tilde r(0) =
v\tilde p(v)\tilde q(u)+up(u)q(0)+r(0)  , 
$$
or
$$
up(u)(q(v)- q(0))+\tilde r(0) =
v\tilde p(v)(\tilde q(u)-\tilde q(0))+r(0)  , 
$$
We note that $r(0) = \tilde r(0)$, because all the other terms in this equation
are divisible by $uv$. That is, we have
$$
up(u)(q(v)- q(0))=
v\tilde p(v)(\tilde q(u)-\tilde q(0)).
$$
Assume first that $q(v)-q(0)$ is not identically zero; that is, 
$q$ is not a constant. The equality just derived allows us to write
(with a suitable ``shift'' of the constants of proportionality).
$$
up(u) = \tilde q(u)-\tilde q(0),\quad\quad\text{and}\quad\quad
v\tilde p(v) = q(v)- q(0).
$$
That is, we have,
\begin{align*}
f(u,v) &= up(u)q(v)+r(v)
 = up(u)q(v)+v\tilde p(v)\tilde q(0)+\tilde r(0)\\
&= (\tilde q(u)-\tilde q(0))q(v)+(q(v)-q(0))\tilde q(0)+r(0)
= \tilde q (u)q(v)+r(0)-\tilde q(0)q(0).
\end{align*}

In the other case, $q(v)$ is a constant $c_0$, so (\ref{weakxy}) yields $f(u,v) = c_0up(u) + r(v)$.

That is, we have shown that $f$ is of one of the forms $\varphi(u)+\psi(v)$ or (up to an additive constant) $\varphi(u)\cdot \psi(v)$, for suitable univariate polynomials $\varphi, \psi$.

Finally, consider the case where $f$ is decomposable. Then we may write $f(u,v)=h(f_0(u,v))$, where $f_0$ is an indecomposable bivariate polynomial over $\R$, and $h$ is a (non-linear) univariate polynomial over $\R$. We let $C_0:=h^{-1}(C)$ denote the pre-image of $C$ under $h$. Note that since $h$ is a polynomial of degree at most $d$ (actually, at most $d/2$), every $c\in C$ has at most $d$ values $c'\in\R$ for which $h(c')=c$. Thus, $|C_0|\le d|C|$, and the number $M_0$ of intersections of the surface $z=f_0(u,v)$ with the point set $A\times B\times C_0$ in $\R^3$ is at least $M$. By the above analysis, applied to the polynomial $f_0$ and to the sets $A$, $B$ and $C_0$, we conclude that either $M=\Theta(M_0)=O(|A|^{2/3}|B|^{1/2}|C|^{2/3}+|A|^{1/2}|B|^{2/3}|C|^{2/3}+|A|+|B|)$, or $f_0$ is of one of the two forms $\varphi(u)+\psi(v)$ or $\varphi(u)\cdot \psi(v)$ (the extra additive term that we got in the latter case can be ``transferred" to the expression defining $h$). Hence, either $f$ is of one of the two forms $h(\varphi(u)+\psi(v))$ or $h(\varphi(u)\cdot \psi(v))$, or $M$ satisfies the bound in Proposition~\ref{prop:part1}, as asserted. This finally concludes the proof of the proposition. $\square$

%---------------------------------------------------------------Proof of Proposition:incid--------------------------------------
\subsection{Proof of Proposition~\ref{incid}}\label{subsec:incid}
We apply Sz\'ekely's technique~\cite{Sz97}, which is based on the Crossing Lemma, as formulated in Theorem~\ref{thm:cross} (see also \cite[p. 231]{PA95}). As noted, this is also the approach used in the proof of Theorem \ref{PS} in Pach and Sharir~\cite{PS98}, but the possible overlap of curves requires some extra (and more explicit) care in the application of the technique.\footnote{The reason why we cannot apply Theorem \ref{PS} directly (with $k=2$) is that it is possible for a pair of points of $\Pi$ to have a non-constant arbitrarily large number of curves that pass through both of them; see the analysis below.}

We begin by constructing a plane embedding of a multigraph $G$, whose vertices are the points of $\Pi$, and each of whose edges connects a pair $\pi_1=(\xi_1,\eta_1)$,
$\pi_2=(\xi_2,\eta_2)$ of points that lie on the same curve $\gamma_{a,c}$ and are consecutive along (some connected component of) $\gamma_{a,c}$; the edge is drawn along the portion of the curve between the points. One edge for each such curve (connecting $\pi_1$ and $\pi_2$) is generated, even when the curves coincide or overlap. Thus there might potentially be many edges of $G$ connecting the same pair of points, whose drawings coincide. Nevertheless, by assumption, the amount of overlap at any specific arc is at most $m_0$.

In spite of this control on the number of mutually overlapping (or, rather, coinciding) edges, we still face the potential problem that the edge multiplicity in $G$ (over all curves, overlapping or not, that connect the same pair of vertices) may not be bounded (by a constant). More concretely, we want to avoid edges $(\pi_1,\pi_2)$ whose multiplicity exceeds $m_0$. (By what has just been argued, not all drawings of such an edge can coincide.)

To handle this situation, we observe that, by the symmetry of the definition of the curves, $\pi_1,\pi_2\in\gamma_{a,c}$ if and only if $(a,c)\in\gamma_{\xi_1,\eta_1}\cap\gamma_{\xi_2,\eta_2}$. Hence, if the multiplicity of the edge connecting $\pi_1$ and $\pi_2$ is larger than $m_0$ then the curves $\gamma_{\xi_1,\eta_1}$ and $\gamma_{\xi_2,\eta_2}$ intersect in more than $m_0$ points, and therefore, since each is the zero set of a polynomial of degree $d$, and since $m_0\ge d^2$, B\'ezout's theorem (Theorem \ref{bezout}) implies that these curves must overlap in a common irreducible component.

Note that, for a given $(\xi_1,\eta_1)$, the curve $\gamma_{\xi_1,\eta_1}$, having degree $d$, has at most $d$ irreducible components, and, by the assumption on $\Gamma$, at most $m_0$ curves share a common irreducible component. That is, each $(\xi_1,\eta_1)$ has at most $(m_0-1)d$ ``problematic'' neighbors that we do not want to connect it to; for any other point, the multiplicity of the edge connecting $(\xi_1,\eta_1)$ with that point is at most $m_0$; more precisely, at most $m_0$ curves $\gamma_{a,c}$ pass through both points.

Consider a point $(\xi_1,\eta_1)$ and one of its bad neighbors $(\xi_2,\eta_2)$; that is, they are points that lie on too many common curves. Let $\gamma_{a,c}$ be one of the curves along which $(\xi_1,\eta_1)$ and $(\xi_2,\eta_2)$ are neighbors.\footnote{We make the pessimistic assumption that they are (consecutive) neighbors along all these curves, which of course does not have to be the case.} Then, rather than connecting $(\xi_1,\eta_1)$ to $(\xi_2,\eta_2)$ along $\gamma_{a,c}$, we continue along the curve from $(\xi_1,\eta_1)$ past $(\xi_2,\eta_2)$ until we reach a good point for $(\xi_1,\eta_1)$, and then connect $(\xi_1,\eta_1)$ to that point (along $\gamma_{a,c}$). We skip over at most $(m_0-1)d$ points in the process, but now, having applied this ``stretching'' to each pair of bad neighbors, each of the modified edges has multiplicity at most $2m_0$ (the factor 2 comes from the fact that a new edge can be obtained by stretching an original edge from either endpoint).

Note that this edge stretching does not always succeed: It will fail when the connected component $\gamma'$ of $\gamma_{a,c}$ along which we connect the points contains fewer than $(m_0-1)d+2$ points of $\Pi$, or when $\gamma'$ is unbounded and there are fewer than $(m_0-1)d$ points of $\Pi$ between $\pi_1,\pi_2$, and an ``end" of $\gamma$. Still, the number of new edges in $G$ is at least $I(\Pi,\Gamma)-\lambda|\Gamma|$, for a suitable constant $\lambda$, where the term $\lambda|\Gamma|$ accounts for missing edges on connected components of the curves, for the reasons just discussed. By what have just been argued, the number of edges lost on any single component is at most $O(m_0d)$, so $\lambda=O(m_0d^2)=O(1)$. 

The final ingredient needed for this technique is an upper bound on the number of crossings between the (new) edges of $G$.  Each such crossing is a crossing between two curves of $\Gamma$.  Even though the two curves might overlap in a common irreducible component (where they have infinitely many intersection points, none of which is a crossing\footnote{In particular, overlapping edges drawn along such a component are not considered to cross one another.}), the number of proper crossings between them is $O(1)$, as follows, for example, from the Milnor--Thom theorem (see \cite{Mil64,Th65}), or B\'ezout's theorem (Theorem \ref{bezout}). Finally, because of the way the drawn edges have been stretched, the edges, even those drawn along the same original curve $\gamma_{a,c}$, may now overlap one another, and then a crossing between two curves may be claimed by more than one pair of edges.  Nevertheless, since no edge straddles through more than $(m_0-1)d$ points, the number of pairs that claim a specific crossing is $O(m_0d)=O(1)$.  Hence, we conclude that the total number of edge crossings in $G$ is $O(|\Gamma|^2)$.

We can now apply Theorem~\ref{thm:cross}, and conclude that
$$
I(\Pi,\Gamma)-\lambda|\Gamma|=O\left(|\Pi|+|\Pi|^{2/3}|\Gamma|^{2/3}\right),
$$
or
$$
I(\Pi,\Gamma) = O\left( |\Pi|^{2/3}|\Gamma|^{2/3} + |\Pi| + |\Gamma| \right) ,
$$
with the constant of proportionality depending on $d$, as asserted. $\square$

%----------------------------------------------Proof of Proposition:weakstruct
\subsection{Proof of Proposition \ref{weakstruct}}\label{subsec:weakstruct}
We may assume that $\gamma'$ does not contain any portion that is contained in a horizontal line, or, since $\gamma'$ is assumed to be irreducible, that $\gamma'$ is not a horizontal line.
%
%For this, first note that it suffices to consider only curves $\gamma_{a,c}$ for which $f(a,v)$ is non-constant (regarded as a polynomial in $v$). Indeed, there are at most $d_u$ values of $a$, for which $f(a,v)$ is independent of $v$. By applying the Schwartz-Zippel lemma (Lemma~\ref{SZ}) to the polynomials $(v,w)\mapsto f(a,v)-w$, for each of these exceptional values $a$, it follows that the number of triples $(a,b,c)\in A\times B\times C$ for which $a$ is problematic (in the above sense) and $f(a,b)=c$ is $d_u\cdot O((|B|+|C|)d_v)$, which is subsumed in the asserted bound on $M$. 
%
%Next, consider a value $z\in\R$ for which $f(u,z)$ is independent of $u$. There are at most $d_v$ such values of $z$, and each of them determines exactly one value $c=f(u,z)$ (which depends only on $z$). By applying the Schwartz-Zippel lemma, this time to the polynomials $(u,v)\mapsto f(u,v)-c$, for each of these at most $d_v$ special values $c$, it follows that the number of triples $(a,b,c)\in A\times B\times C$, for which $c$ is one of these exceptional values and $f(a,b)=c$, is $d_v\cdot O((|A|+|B|)d)$. This is again subsumed in the asserted bound on $M$, and we may ignore such values $c\in C$ in our analysis. To recap, by trimming $A$ and $C$ accordingly, we may now assume that (i) for each $a\in A$, $f(a,z)$ is non-constant in $z$, and (ii) for each $c\in C$, no value $z_0\in\R$ yields a constant polynomial $f(u,z_0)$ (i.e., independent of $u$) whose value is $c$.   
%
Indeed, if $\gamma'$ were a horizontal line of the form $y=\eta_0$ then, for any curve $\gamma_{a,c}$ that contains $\gamma'$, the system
\begin{align*}
f(a,z)&=\eta_0\\
f(\xi,z)&=c,
\end{align*}
in the variables $\xi,z$, would have infinitely many solutions. By our assumption, made at the beginning of the analysis, $f(a,z)$ is non-constant in the variable $z$, and hence $z$ is one of the at most $d_v$ roots of  $f(a,z)=\eta_0$. Hence, to get infinitely many solutions $\xi,z$, it must be that $\xi\mapsto f(\xi,z)$ is independent of $\xi$. But then $z$ is one of the exceptional values discussed at the beginning of the analysis, and our pruning of $C$ ensures that $f(\xi,z)\not\in C$, and hence $f(\xi,z)=c$ does not have infinitely many solutions. That is, in the reduced configuration, $\gamma'$ cannot be a horizontal line.

Let $(\xi,\eta)$ be a regular point of $\gamma'$, and let $(\alpha,\beta)$ denote the direction vector of the line tangent to $\gamma'$ at $(\xi,\eta)$. For reasons to be clarified later, we choose, as we may, the point $(\xi,\eta)$ so that $f_u(\xi,v)$ (regarded as a polynomial in $\R[v]$) is non-constant, and so that the polynomial $\eta-f(u,v)\in\R[u,v]$ is irreducible over $\R$. Indeed, for the former property we only need to avoid the at most $d_u$ (common) zeros $\xi$ of the coefficients of the nonconstant monomials of $f_u$ (regarding $f_u$ as a polynomial in $v$). For the latter property, we use our assumption that $f$ is indecomposable. In this case Corollary \ref{cordecom} says that there are at most a constant number of values $\eta$ for which $\eta-f(u,v)$ is reducible over $\R$. Hence, in total, since $\gamma'$ does not contain any horizontal line, we need to avoid at most a constant number of points $(\xi,\eta)$ on $\gamma'$ to have these two properties, and we let $(\xi,\eta)$ be one of the other (infinitely many) regular points of $\gamma'$.

By assumption, there are $m>d_ud_v+d_ud(d+d_v-1)$ pairs $(a_i,c_i)$, $i=1,\ldots,m$, such that the curves $\gamma_{a_i,c_i}$ all contain $\gamma'$ (and in particular, a neighborhood of $(\xi,\eta)$ along $\gamma'$). We recall the definitions, for the convenience of the reader.
$$
\bar\gamma_{a_i,c_i}:=\{(x,y,z)\in\R^3\mid y=f(a_i,z)\land c_i=f(x,z)\}=\sigma_{a_i}\cap\sigma_{c_i}^*,
$$
where
$$
\sigma_{a_i}:=\{(x,y,z)\in\R^3\mid y=f(a_i,z)\}
$$
$$
\sigma_{c_i}^*:=\{(x,y,z)\in\R^3\mid c_i=f(x,z)\},
$$
and
$$\gamma_{a_i,c_i}:=\{(x,y)\in\R^2\mid \exists z\in \R~\text{such that}~y=f(a_i,z)\land c_i=f(x,z)\}.
$$
Then, for each $i=1,\ldots,m$, there exists a point $z_i\in\R$ for which $p_i:=(\xi,\eta,z_i)\in\bar\gamma_{a_i,c_i}$. Observe that the values $z_i$, $i=1,\ldots,m$, are not necessarily distinct, but nevertheless the cardinality $m'$ of $\{z_i\mid i=1,\ldots,m\}$ is at least $m/d_u$. Indeed, for a given value $z_0$, the equation $c=f(\xi,z_0)$ determines $c$ uniquely, and the equation $\eta=f(a,z_0)$ determines at most $d_u$ possible values of $a$. (For the latter claim, we note that $f(a,z_0)$ cannot be independent of $a$ and satisfy $f(a,z_0)\equiv \eta$, for that would imply that $\eta-f(u,v)$ is divisble by $v-z_0$, contradicting the assumption that $\eta-f(u,v)$ is irreducible and that $d\ge 2$.) Then there are at most $d_u$ pairs $(a_i,c_i)$ with $(\xi,\eta,z_0)\in\bar\gamma_{a_i,c_i}$, and hence at most $d_u$ indices $i$ for which $z_i=z_0$. Thus  $|\{z_i\mid i=1,\ldots,m\}|\ge m/d_u$, as claimed. Therefore we may assume, by re-indexing if needed, that the values $z_1,\ldots,z_{m'}$ are distinct, with $m'>d_v+d(d+d_v-1)$. 

Observe that for at least $m'-d_v$ of the indices $1\le i\le m'$, the point $p_i$ is a regular point of both $\sigma_{a_i},\sigma_{c_i}^*$. Indeed, $\sigma_{a_i}$ is a smooth surface since it is the (cylindrical) graph of the polynomial function $y=f(a_i,z)$. $p_i$ is singular in $\sigma_{c_i}^*$ only if $f_u(\xi,z_i)=0$, but this equation is satisfied by at most $d_v$ values $z_i$. (Recall that by our choice of $\xi$, the polynomial $f_u(\xi,z)$ is non-constant in $z$.) Therefore we may assume, by re-indexing if needed, that each of the points $p_1,\ldots,p_{m''}$ is regular on both $\sigma_{a_i}$ and $\sigma_{c_i}^*$, with $m''>d(d+d_v-1)$. 

Let $\pi_{a_i},\pi_{c_i}^*$ be the tangent planes of $\sigma_{a_i},\sigma_{c_i}^*$ at $p_i$, which are now well defined. The normal vectors of $\pi_{a_i},\pi_{c_i}^*$ at $(\xi,\eta,z_i)$ are 
$$
{\bf n}_{a_i}=(0,1,-f_v(a_i,z_i)),\quad{\bf n}_{c_i}^*=(f_u(\xi,z_i),0,f_v(\xi,z_i)),
$$
respectively. Note that these values imply that $\pi_{a_i}\neq \pi_{c_i}^*$, and thus the intersection $\pi_{a_i}\cap\pi_{c_i}^*$ is a line $l$. The direction vector of $l$ is orthogonal to both ${\bf n}_{a_i},{\bf n}_{c_i}^*$, and is thus given by
$$
{\bf n}_{a_i}\times {\bf n}_{c_i}^*=
\left|
\begin{array}{ccc}
{\bf i}& {\bf j}& {\bf k}\\
{\bf n}_{a_i,x}& {\bf n}_{a_i,y} & {\bf n}_{a_i,z}\\
{\bf n}_{c_i,x}^*&{\bf n}_{c_i,y}^*&{\bf n}_{c_i,z}^*
\end{array}\right|
=
\left|
\begin{array}{ccc}
{\bf i}& {\bf j}& {\bf k}\\
0& 1 & -f_v(a_i,z_i)\\
f_u(\xi,z_i)&0&f_v(\xi,z_i)
\end{array}\right|
=
\left(
\begin{array}{c}
f_v(\xi,z_i)\\
-f_v(a_i,z_i)f_u(\xi,z_i)\\
-f_u(\xi,z_i)
\end{array}\right).
$$

By the assumption on $(\xi,\eta)$, the projection of ${\bf n}_{a_i}\times {\bf n}_{c_i}^*$ onto the $xy$-plane is parallel to $(\alpha,\beta)$ (recall that $(\alpha,\beta)$ depends only on $\gamma'$ and not on a specific choice of $(a_i,c_i)$). That is, for every $i=1,\ldots,m''$, we have
$$
\beta f_v(\xi,z_i)+\alpha f_v(a_i,z_i)f_u(\xi,z_i)=0.
$$

Consider the system of equations
\begin{equation}\label{1}
\left\{
\begin{array}{l}
g_1(a,z):=\beta f_v(\xi,z)+\alpha f_v(a,z)f_u(\xi,z)=0\\
g_2(a,z):=\eta-f(a,z)=0,
\end{array}\right.
\end{equation}
with $a,z$ being the unknowns. That is, (\ref{1}) is satisfied by the $m''>d(d+d_v-1)$ distinct pairs $(a_i,z_i)$, $i=1,\ldots,m''$, so it has at least $d(d+d_v-1)+1$ solutions. Since $\deg g_1\le d+d_v-1$ and $\deg g_2=d$, B\'ezout's theorem (Theorem \ref{bezout}) implies that the polynomials $g_1(a,z)$ and $g_2(a,z)$ in $\R[a,z]$ must have a (non-constant) common factor. Recalling that, by our choice of $\eta$, $g_2$ is irreducible over $\R$, it follows that $g_2$ divides $g_1$.

Note that the variable $a$ has the same degree in both $g_1$ and $g_2$. Indeed, its degree in $g_2$ is $d_u$ and its degree in $g_1$ is at most $d_u$; if the latter degree were smaller than $d_u$, $g_2$ could not divide $g_1$. Hence $g_1$ must be of the form
\begin{equation}\label{2}
g_1(a,z)\equiv g_2(a,z)h(z),
\end{equation}
with $h$ being independent of $a$. We write
$$
f(u,v)=\sum_{k=0}^{d_u}c_k(v)u^k,
$$
so
$
f_v(u,v)=\sum_{k=0}^{d_u} c_k'(v)u^k.
$
Then (\ref{2}) becomes
$$
\beta f_v(\xi,z)+\alpha \left(\sum_{k=0}^{d_u} c_k'(z)a^k\right)f_u(\xi,z)\equiv \left(\eta-\sum_{k=0}^{d_u} c_k(z)a^k\right)h(z),
$$
or
$$
\beta f_v(\xi,z)+ \alpha f_u(\xi,z)c_0'(z)+\sum_{k=1}^{d_u} \alpha f_u(\xi,z)c_k'(z)a^k \equiv \eta h(z)-h(z)c_0(z)-\sum_{k=1}^{d_u}h(z)c_k(z)a^k.
$$
Hence we have, in particular,
$$
c_k'(z)\equiv-\frac{h(z)}{\alpha f_u(\xi,z)}c_k(z),
$$
for every $1\le k\le d_u$. Hence, for every pair of distinct indices $1\le k,l\le d_u$, we have
$$
c_k'(z)c_l(z)\equiv c_k(z)c_l'(z),
$$
or, when $c_l(z)$ not identically zero, 
$
\left(\frac{c_k(z)}{c_l(z)}\right)'\equiv 0,
$
or
$
\frac{c_k(z)}{c_l(z)}\equiv \beta_{kl},
$
for $\beta_{kl}$ a constant.
That is, there exist constants $\lambda_k\in \R$ and a polynomial $q(z)$ (independent of $k$), such that
$
c_k(z)\equiv \lambda_kq(z),
$
for $k=1,\ldots,d_u$. (This also takes care of coefficients $c_k(z)$ that are identically zero.)
Hence,
$$
f(u,v)=c_0(v)+ uq(v)\sum_{k=0}^{d_u-1}\lambda_{k+1}u^{k}.
$$
Putting $p(u):=\sum_{k=0}^{d_u-1}\lambda_{k+1}u^{k}$, we conclude that $f$ is of the form $f(u,v)=c_0(v)+ up(u)q(v)$, as asserted. $\square$

%-----------------------------------------------------PART TWO--------------------------------------------------------------------
\section{Proof of Theorem \ref{main2}: Part 2}\label{sec:part2}
%--------------------------------------------------------------------------------------------------------------------------------------

So far we have considered two reductions where the parametric plane in which point-curve incidences have been analyzed contained $A\times C$ and $B\times C$, respectively. In this section we consider a somewhat more natural (or ``standard") setup, in which the dependence on $z$ is eliminated right away, and the emphasis is mainly on the sets $A$ and $B$. This approach leads to the second upper bound in Theorem \ref{main2}. That is, we show:
%---------prop:Part2-------------------
\begin{prop}\label{prop:part2}
Let $A$, $B$, and $C$ be three finite sets of real numbers. Let $f\in\R[u,v]$ be a bivariate real polynomial of constant degree $d\ge 2$, and let $M$ denote the number of intersection points of the surface $w=f(u,v)$ with $A\times B\times C$ in $\R^3$. Then either 
$$
M= O\left(\;\big(|A|^{2/3}|B|^{2/3}+|A|+|B|\big)|C|^{1/2}\;\right),
$$
with a constant of proportionality that depends on $d$, or $f$ is of one of the forms $f(u,v)=h(\varphi(u)+\psi(v))$, or $f(u,v)=h(\varphi(u)\cdot\psi(v))$, for some real univariate polynomials $\varphi,\psi,h$.
\end{prop}

\noindent{\bf Proof.}
Arguing as in Section~\ref{sec:part1}, we may assume, without loss of generality, that $f$ is indecomposable. In the present ``standard'' setup, the curves are defined by
\begin{equation}\label{curvpri}
\gamma_{a,b} = \{ (x,y) \mid f(a,x) = f(b,y) \} ,
\end{equation}
for $a,b\in A$. We let $\Gamma$ denote the multiset of these curves (so $|\Gamma|=|A|^2$, counting curves with multiplicity), and put $\Pi:= B^2$. For each $c\in C$ let $M_c$ denote the number of pairs $(a,b)\in A\times B$ such that $f(a,b)=c$. Then $M=\sum_{c\in C}M_c$. Let $I=I(\Pi,\Gamma)$ denote the number of incidences between the points of $\Pi$ and the curves of $\Gamma$. A similar argument to that in the preceding section shows that, for every pair of pairs $(a_1,b_1),(a_2,b_2)\in A\times B$ with $f(a_1,b_1)=f(a_2,b_2)$, we have $(b_1,b_2)\in\gamma_{a_1,a_2}$. This implies, as before, that $I\ge \sum_{c\in C}M_c^2$, and then
\begin{equation}\label{M}
M=\sum_{c\in C} M_c\le \left(\sum_{c\in C}M_c^2\right)^{1/2}\cdot|C|^{1/2}\le I^{1/2}|C|^{1/2}.
\end{equation}
Hence the problem is reduced to obtaining an upper bound on $I$. 

%%%
\paragraph{Bounding the number of incidences.}
%%%
Again, we are concerned with situations where many curves of $\Gamma$ share a common irreducible component $\gamma'$. We want to show that in this case $f$ must have one of the special forms asserted in the theorem. The complementary case, in which no component $\gamma'$ is shared by too many curves, will lead to the incidence bound that we are after, as we detail next.

Note that in the definition of the curves (\ref{curvpri}), the roles of the variables $u,v$ of $f$ are symmetric and can be reversed. Namely, we can consider the ``dual" curves\footnote{It is interesting to observe that, in contrast, in the setup of Section~\ref{sec:part1} the dual scenario involves the same kind of curves as the primal one.}
\begin{equation}\label{curvdu}
\gamma_{\xi,\eta}^* = \{ (x,y) \mid f(x,\xi) = f(y,\eta) \},
\end{equation}
for $\xi,\eta\in B$. Then, for $(a_1,b_1),(a_2,b_2)\in A\times B$, we have that $(b_1,b_2)\in\gamma_{a_1,a_2}$ if and only if $(a_1,a_2)\in\gamma_{b_1,b_2}^*$. We let $\Gamma^*$ denote the multiset of the curves $\gamma_{\xi,\eta}^*$, with $(\xi,\eta)\in B^2$, and put $\Pi^*:=A^2$. 

Let $\Gamma_0$ (resp., $\Gamma_0^*)$ denote the set of irreducible curves that are shared by more than $m_0:=\max\{d_u^2,d_v^2\}$ curves of $\Gamma$ (resp., of $\Gamma^*$). 
Incidences between points $(\xi,\eta)\in \Pi$ and curves $\gamma_{a,b} \in\Gamma$, such that the portion of $\gamma_{a,b}$ containing $(\xi,\eta)$ is not in $\Gamma_0$, and the portion of $\gamma_{\xi,\eta}^*$ containing $(a,b)$ is not in $\Gamma_0^*$, can be analyzed via Sz\'ekely's crossing-lemma technique (see Theorem~\ref{thm:cross}), as we did in the proof of Proposition~\ref{incid} in Section \ref{sec:part1}, and their number is thus at most
\begin{equation} \label{goodinc}
O\left( |\Pi|^{2/3}|\Gamma|^{2/3} + |\Pi| + |\Gamma| \right) =
O\left( |A|^{4/3}|B|^{4/3} + |A|^2 + |B|^2 \right),
\end{equation}
where the constant of proportionality depends on $d$. (In more details, the only difference from the proof of Proposition~\ref{incid} is in analyzing the multiplicity of the edges in the constructed graph $G$; this multiplicity can be interpreted as the number of dual curves that share an irreducible factor, and hence is bounded by $m_0$, due to our exclusion of incidences that occur on curves of $\Gamma_0^*$.)

If at least half the quadruples in $Q:=\{(a,\xi,b,\eta)\in (A\times B)^2\mid f(a,\xi)=f(b,\eta)\}$ correspond to incidences that occur on primal curves that are not in $\Gamma_0$, and on dual curves that are not in $\Gamma_0^*$, then the expression in (\ref{goodinc}) serves as the desired upper bound on $I$, which, combined with (\ref{M}), yields the bound on $M$ asserted in the proposition. We may thus assume that at least half the quadruples in $Q$ correspond to incidences that occur either on primal curves of $\Gamma_0$ or on dual curves of $\Gamma_0^*$, and, without loss of generality (by the symmetry of the two settings), that at least a quarter of the quadruples in $Q$ correspond to incidences that occur on primal curves of $\Gamma_0$. We refer to the curves in $\Gamma_0$ as \emph{popular curves}.

%-------------------Curves with larger multiplicity------------------------------------
\paragraph{Curves with larger multiplicity: Reducibility and its consequences.} 

Consider then incidences that occur on popular curves. Let $Q_0$ denote the subset of quadruples $(a,p,b,q)$ in $Q$ such that the irreducible component of $\gamma_{a,b}$ that contains $(p,q)$ is popular. We have the following key proposition, whose proof is given in Section~\ref{sec:fh}.
%--------------------------------Proposition:fh-----------------------------------------
\begin{prop} \label{prop:fh}
There exists a bivariate polynomial $h$, that depends only on $f$, such that $\deg h\le 2d^2$, and $h$ is of one of the forms $\varphi(u)+\psi(v)$, or $\varphi(u)\cdot\psi(v)$, for some univariate polynomials $\varphi,\psi$ (that depend only on $f$), and such that the following property holds. For at least $|Q_0|-O(|A||B|)$ quadruples $(a,p,b,q)$ of $Q_0$, we have $h(a,p)=h(b,q)$.
\end{prop}

Recall that we are currently assuming that at least a quarter of the quadruples in $Q$ correspond to incidences that occur on curves of $\Gamma_0$. That is, $|Q_0|=\Theta(|Q|)$. We remove from $C$ all elements $c$ for which $c-f(u,v)$ is reducible. Since we continue to assume that $f$ is indecomposable, Corollary~\ref{cordecom} tells us that the number of values $c$ for which $c-f(u,v)$ is reducible is smaller than $d$, and an application of the Schwartz-Zippel lemma (Lemma~\ref{SZ}) implies that, for each such $c$, the number of pairs $(a,p)\in A\times B$ satisfying $f(a,p)=c$ is $O((|A|+|B|)d)$, so the number of quadruples associated with these exceptional values is $O((|A|+|B|)^2)$, and we simply ignore them in our analysis (recall that this argument has already been used in earlier parts of the analysis). We may assume that $|Q_0|$ is much larger than this bound, for otherwise $|Q|$ satisfies the bound in (\ref{goodinc}) and we are done. By removing at most $O(|A||B|)$ additional quadruples from $Q_0$, as prescribed in Proposition~\ref{prop:fh}, we may also assume that $h(a,p)=h(b,q)$, for each surviving quadruple $(a,p,b,q)\in Q_0$.

With these reductions, we conclude that there exists a pair $(a,p)\in A\times B$ that participates in at least $t:=|Q_0|/(|A||B|)$ quadruples $(a,p,b,q)$ of $Q_0$, so that, for $c=f(a,p)=f(b,q)$, $c-f(u,v)$ is irreducible, and $c'=h(a,p)=h(b,q)$. We put $c'=h(a,p)$, and note that $c$ and $c'$ are fixed once $(a,p)$ is fixed.

Assume first that the polynomial $h(u,v)$ is indecomposable. In this case we can apply a fully symmetric argument to $h$, and, by possibly discarding another set of $O((|A|+|B|)^2)$ quadruples from $Q_0$, involving pairs $(a,p)$ for which $h(a,p)-h(u,v)$ is reducible, be left with quadruples involving only pairs $(a,p)$ for which $(a,p)$ satisfies all the properties assumed so far, and $h(a,p)-h(u,v)$ is also irreducible. 

Now fix a pair $(a,p)\in A\times B$ satisfying all the above properties. For at least $t$ pairs $(b,q)$, we have
\begin{align*}
& f(a,p) - f(b,q) = 0 \\
& h(a,p) - h(b,q) = 0 .
\end{align*}
That is, the polynomials $f(a,p) - f(u,v)$ and $h(a,p) - h(u,v)$ have at least $t$ common roots. Hence, unless $t$ is at most some suitable constant (in which case $|Q|=O(|Q_0|)=O(|A||B|)$, well below the bound in Proposition~\ref{prop:part2}), these polynomials have a common factor. But since $f(a,p)-f(u,v)$ and $h(a,p)-h(u,v)$ are both irreducible, they must be proportional to one another, implying that
$$
f(u,v) = \alpha h(u,v) + \beta ,
$$
for suitable constants $\alpha$, $\beta$. 

Next we consider the case where $h$ is decomposable. We have the following simple claim. 
%------------Claim:sum--------------------
\begin{clm}\label{clm:sum}
Let $h$ be any polynomial of the form $h(u,v)=\varphi(u)+\psi(v)$, for some non-constant univariate real polynomials $\varphi,\psi$. Then $h$ is indecomposable.
\end{clm}
\noindent{\bf Proof.}
Suppose to the contrary that $h$ is decomposable, and write $h(u,v) = r(h_0(u,v))$, for some nonlinear univariate polynomial $r$ and a bivariate polynomial $h_0$. We have
$$
\varphi(u)+\psi(v)=r(h_0(u,v)).
$$
Taking derivatives of both sides with respect to the variable $u$ yields
$$
\varphi'(u)=r'(h_0(u,v))(h_0)_u(u,v).
$$
Since by assumption $\varphi'(u)\neq 0$ (and by the unique factorization property over $\R$), $r'(h_0(u,v))$ must divide $\varphi'(u)$ and thus be independent of $v$. Since $r$ is nonlinear, this is easily seen to imply that $h_0$ itself, and thus $h$ too, must be independent of $v$, contradicting the assumption that $\psi(v)$ is non-constant.
$\Box$

Hence, a decomposable $h$ must be of the form $h(u,v)=\varphi(u)\psi(v)$. We write $h(u,v)=r(h_0(u,v))$, where $h_0$ is indecomposable, and $r$ is a nonlinear univariate real polynomial. Using the same argument as before, we may assume that we are left with quadruples involving only pairs $(a,p)$ for which all the previous properties are satisfied (with the values $c,c'$ fixed), and $h_0(a,p)-h_0(u,v)$ is irreducible. Now consider the equation $r(s)=c'$, which has at most $\deg h\le d^2$ real roots, and enumerate those roots $s$ for which $s-h_0(u,v)$ is irreducible as $s_1,\ldots,s_{d'}$, for some $d'\le d^2$. We have at least $t$ elements $(a,p)\in (A\times B)_c$ for which $h_0(a,p)\in \{s_1,\ldots,s_{d'}\}$, and so there is an index $j$ such that at least $t/d'\ge t/d^2$ elements $(a,p)\in (A\times B)_c$ satisfy $h_0(a,p)=s_j$. 

Now fix $(a,p)\in (A\times B)_c$ as one of these pairs. Then, for at least $t/d^2$ pairs $(b,q)$, which share the same properties with $(a,p)$, we have
\begin{align*}
& f(a,p) - f(b,q) = 0 \\
& h_0(a,p) - h_0(b,q) = 0 .
\end{align*}
That is, the two irreducible polynomials $f(a,p) - f(u,v)$ and $h_0(a,p) - h_0(u,v)$ have at least $t/d^2$ common roots. Arguing as in the previous case, assuming $t$ to be sufficiently large, this implies that
$$
f(u,v) = \alpha h_0(u,v) + \beta ,
$$
for suitable constants $\alpha$, $\beta$. 

To complete the analysis we claim that $h_0$ is of the form $h_0(u,v)= \varphi_1(u)\psi_1(v)+z$, for some real univariate polynomials $\varphi_1,\psi_1$, and $z\in\R$. Indeed, we can factor $r$ over $\R$ into a product of linear and irreducible quadratic factors. If there is at least one linear factor then the corresponding factor of $r(h_0(u,v))$, of the form $h_0(u,v)-z$, for some $z\in\R$, divides $\varphi(u)\psi(v)$, and thus must be of the form $\varphi_1(u)\psi_1(v)$, implying the claim. Otherwise, consider an irreducible factor of the form $(h_0-w)^2+z^2$, for $z,w\in\R$. That is, we must have
$$
(h_0(u,v)-w)^2+z^2=\varphi_1(u)\psi_1(v),
$$
for suitable polynomials $\varphi_1,\psi_1$. Taking derivatives of both sides of this identity with respect to the variable $u$, we get
$$
2(h_0(u,v)-w)(h_0)_u(u,v)=\varphi_1'(u)\psi_1(v).
$$
Hence $h_0(u,v)-w$ must divide $\varphi_1'(u)\psi_1(v)$, so, as above, $h_0$ has the asserted form.

In summary, we have covered all subcases, and have shown that the existence of a large number of curves that overlap in a common irreducible component implies that $f$ has one of the special forms. So far we have assumed that $f$ is indecomposable, but, as noted in the beginning of the proof, the case where $f$ is decomposable can be handled as in Section~\ref{sec:part1}. We have thus completed the proof of Proposition~\ref{prop:part2}, that is, of the second part of the proof of Theorem~\ref{main2}. 
$\Box$

%-----------------------------------Proof of Proposition prop:fh-------------------------------------
\subsection{Proof of Proposition~\ref{prop:fh}}\label{sec:fh}
We begin the analysis with the following lemma, which derives a useful property involving reducibility of bivariate polynomials of the special form $p(x)-q(y)$ that we consider here. To prepare for the lemma, we introduce the following notation. For a bivariate polynomial $u(x,y)$, let $u^*(x,y)$ denote the polynomial which is the sum of all monomials of $u(x,y)$ of maximum total degree. We refer to $u^*(x,y)$ as the \emph{leading-terms polynomial}, or \emph{LT-polynomial} in short, of $u(x,y)$. Note that if $u(x,y)=v(x,y)w(x,y)$ then necessarily we also have $u^*(x,y)=v^*(x,y)w^*(x,y)$. 
\begin{lem} \label{common}
Let $f,g\in\reals[x,y]$ be two polynomials of the special form
\begin{align*}
f(x,y)&=p_1(x)-q_1(y)\\
g(x,y)&=p_2(x)-q_2(y) ,
\end{align*}
and assume that they have a nontrivial common factor. Assume also that, for $i=1,2$, $\deg p_i = \deg q_i$, and denote this common value by $d_i$. Write
\begin{align*}
[p_1(x)-q_1(y)]^* & = a_1x^{d_1} - b_1y^{d_1} \\
[p_2(x)-q_2(y)]^* & = a_2x^{d_2} - b_2y^{d_2} ,
\end{align*}
for suitable nonzero coefficients $a_1$, $b_1$, $a_2$, $b_2$.
Then we have
\begin{equation} \label{powers}
\left( \frac{a_1}{b_1} \right)^{d_2} =
\left( \frac{a_2}{b_2} \right)^{d_1} .
\end{equation}
\end{lem}
\noindent{\bf Proof.}
By dividing $p_1$ and $q_1$ by $b_1$
and $p_2$ and $q_2$ by $b_2$, we may write
\begin{align*}
f^*(x,y) = [p_1(x) - q_1(y)]^* & = c_1 x^{d_1} - y^{d_1} \\
g^*(x,y) = [p_2(x) - q_2(y)]^* & = c_2 x^{d_2} - y^{d_2} ,
\end{align*}
where $c_1=a_1/b_1\ne 0$ and $c_2=a_2/b_2\ne 0$.

As noted above, the fact that $f$ and $g$ share a common factor implies that $f^*$ and $g^*$ also share a common factor. Hence, the system of equations 
\begin{align*}
c_1 x^{d_1} - y^{d_1} &=0\\
c_2 x^{d_2} - y^{d_2} &=0,
\end{align*}
has infinitely many solutions $(x,y)\in \C\times \C$.
In particular, there exists $(x_0,y_0)\in \C\times \C$, with $x_0\neq 0$ (the above system has a unique solution with $x_0=0$), such that
$$
c_1 =\left(\frac{y_0}{x_0}\right)^{d_1}\quad\quad\text{and}\quad\quad
c_2 =\left(\frac{y_0}{x_0}\right)^{d_2}.
$$
This implies
$$
c_1^{d_2}=c_2^{d_1},\quad\quad\text{or}\quad\quad
\left( \frac{a_1}{b_1} \right)^{d_2} =
\left( \frac{a_2}{b_2} \right)^{d_1},
$$
as asserted. $\Box$

Let us return to the setup under consideration, where we have a multiset $\Gamma$ of curves of the form
$$
\gamma_{a,b} = \{ (x,y) \mid f(a,x) = f(b,y) \} ,
$$
for $a,b\in A$, and we want to analyze the situation where we have an irreducible component $\gamma'$ that is contained in at least $m_0+1=\max \{d_u^2,d_v^2\}+1$ of these curves. Denote by $S(\gamma')$ the set of all the pairs $(a,b)\in A^2$ that define these curves. 

Fix three generic, regular points $(\xi_1,\eta_1),(\xi_2,\eta_2),(\xi_3,\eta_3)\in\gamma'$, so that they satisfy conditions (i) and (ii), spelled out shortly below. We have
\begin{equation}\label{sysp}
f(a,\xi_i) - f(b,\eta_i)= 0,\quad\text{for}~i=1,2,3,~\text{and for}~(a,b)\in S(\gamma'). 
\end{equation}

Write $f(u,v) = \sum_{i=0}^{d_u} c_i(v) u^i$, and let $0\le k\le d_u$ denote the maximal index for which $c_k(v)$ is non-constant in $v$. If $k$ does not exist then $f(u,v)$ does not depend on $u$, so it has (a degenerate version of) one of the special forms in the theorem. If $k=0$ then $f(u,v)$ is of the form $\varphi(u)+c_0(v)$, for a suitable univariate polynomial $\varphi$, so again it has one of the special forms. In both cases $h(u,v):=f(u,v)$ clearly satisfies the property asserted in the proposition.
 
Assume then that $k\ge 1$ and that the points $(\xi_i,\eta_i)$, $i=1,2,3$, are chosen so that (i) $c_k$ does not
vanish at any of the six points $\xi_i$, $\eta_i$, and (ii) for any $1\le i< j\le 3$, $c_k(\xi_i)\neq c_k(\xi_j)$ and $c_k(\eta_i)\neq c_k(\eta_j)$. To see that such a choice is possible, we note that $c_k$ is non-constant and that $\gamma'$ is not a line parallel to one of the axes. (The latter property holds because the polynomial defining $\gamma'$ is a factor of $f(a,x)-f(b,y)$, and we may assume (arguing as in Section~\ref{sec:part1}) that $(a,b)$ is not one of the at most $d_u^2$ pairs for which $f(a,x)-f(b,y)$ is independent of one of the variables $x,y$). Hence, there are infinitely many ways to choose such points $(\xi_i,\eta_i)$. From (\ref{sysp}), we get the system of equations
\begin{align}\label{sysp2}
f(a,\xi_1)-f(a,\xi_2) - f(b,\eta_1)+f(b,\eta_2)&= 0\\
f(a,\xi_3) - f(b,\eta_3)&= 0 ,\nonumber
\end{align}
which, as equations in $a$ and $b$, have at least $m_0+1>d_u^2$ common roots; the purpose of the first equation in~(\ref{sysp2}) is to get rid of the $v$-independent leading $u$-terms of $f$.
Since the first equation in (\ref{sysp2}) has degree $k$ and the second has degree $d_u$, and since $m_0+1>d_u^2\ge d_uk$, B\'ezout's theorem (Theorem \ref{bezout}) tells us that 
$$
f(a,\xi_1)-f(a,\xi_2) - f(b,\eta_1)+f(b,\eta_2),\quad f(a,\xi_3) - f(b,\eta_3)
$$
(regarded as polynomials in $a,b$), have a common factor. We can therefore apply Lemma~\ref{common} to these two polynomials. 
We have
\begin{align*}
[f(a,\xi_1)-f(a,\xi_2) - f(b,\eta_1)+f(b,\eta_2)]^* & = (c_k(\xi_1)-c_k(\xi_2))a^k - (c_k(\eta_1)-c_k(\eta_2))b^k \\
[f(a,\xi_3) - f(b,\eta_3)]^* & = c_{d_u}(\xi_3)a^{d_u} - c_{d_u}(\eta_3)b^{d_u} ,
\end{align*}
where, by construction, all the coefficients in the right-hand sides are non-zero. Lemma~\ref{common} thus implies
that
\begin{equation}\label{eqco}
\left( \frac{ c_k(\xi_1)-c_k(\xi_2) }{ c_k(\eta_1)-c_k(\eta_2) }\right)^{d_u} =
\left( \frac{ c_{d_u}(\xi_3) }{ c_{d_u}(\eta_3) }\right)^k .
\end{equation}

We distinguish between two cases, depending on whether (\ref{eqco}) holds for $k<d_u$ or $k=d_u$.
%----------------------Case: k<d_u----------------------
\paragraph{The case $k<d_u$.} 
Recall our assumption that $k\ge 1$, so we have $1\le k<d_u$. In this case the right hand-side of~(\ref{eqco}) is 1. Hence, replacing $(\xi_1,\eta_1)$ by a generic point\footnote{Note that we may assume that none of the points of $\gamma'\cap B^2$ under consideration is an isolated point of $\gamma'$. Indeed, each curve of $\Gamma$ may have at most $d^2$ such points, and hence the number of incidences between the points of $\Pi$ and the isolated points of curves of $\Gamma$, counted with multiplicity, is at most $d^2|A|^2$.} $(\xi,\eta)$ along $\gamma'$, we conclude that, for all points $(\xi,\eta)\in \gamma'$,
\begin{equation} \label{la-primal2}
c_k(\eta) -c_k(\xi)= \lambda ,
\end{equation}
for some fixed parameter $\lambda$, which depends on $\gamma'$. 

In other words, all points on $\gamma'$ satisfy the system
\begin{align*}
f(a,x) - f(b,y) & = 0 \\
c_k(x)-c_k(y) -\lambda & = 0 , 
\end{align*}
for any of the pairs $(a,b)\in S(\gamma')$. By B\'ezout's theorem, $f(a,x)-f(b,y)$ and $c_k(x)-c_k(y)-\lambda$ have
a common factor.

By Claim~\ref{clm:sum}, $q(x,y):= c_k(x)-c_k(y)$ is indecomposable. Stein's theorem (see Corollary~\ref{cordecom}) then implies that $c_k(x)-c_k(y)-\lambda$ is irreducible for all but fewer than $d_v$ values of $\lambda$. Hence there are fewer than $d_v$ curves $\gamma'$, such that $\gamma'$ is a component of $c_k(x)-c_k(y)-\lambda=0$, for some constant parameter $\lambda$, and $c_k(x)-c_k(y)-\lambda$ is reducible. Thus we can ignore such curves in our analysis, since their (total) contribution to the number of incidences, and hence to the cardinality of $Q$, is $O(|A||B|)$, which is subsumed in the bound of~(\ref{goodinc}). Let $Q_0$ denote the subset of $Q$ obtained by discarding the $O(|A||B|)$ quadruples that correspond to these incidences. We may therefore assume that, for the value of $\lambda$ associated with $\gamma'$, $c_k(x)-c_k(y)-\lambda$ is indeed irreducible. It follows that $c_k(x)-c_k(y)-\lambda$ divides $f(a,x)-f(b,y)$, for every pair $(a,b)\in S(\gamma')$. That is, we may write
\begin{equation}\label{eq:idxy}
f(a,x)-f(b,y) \equiv (c_k(x)-c_k(y)-\lambda) g(x,y),
\end{equation}
for a suitable polynomial $g$ that depends on $a$ and $b$. 

Consider now the symmetric representation $f(u,v) = \sum_{j=0}^{d_v} \tilde c_j(u) v^j$, and let $1\le \ell\le d_v$ denote the maximal index for which $\tilde c_\ell(u)$ is non-constant in $u$. (As before, we may assume that $\ell$ exists and that $\ell\ge 1$.) 

Fix a pair $(a,b)\in S(\gamma')$, and substitute $y=x$ in (\ref{eq:idxy}). By discarding at most $O(|A||B|)$ quadruples of $Q_0$, we may assume that $\tilde c_\ell(a)\neq\tilde c_\ell(b)$. Indeed, by the Schwartz-Zippel lemma (Lemma~\ref{SZ}), the equation $\tilde c_\ell(x)-\tilde c_\ell(y)=0$ has at most $O(|A|)$ solutions $(a,b)\in A^2$. Then, by another application of the Schwartz-Zippel lemma, for each such $(a,b)$, $\gamma_{a,b}$ is incident to at most $O(|B|)$ points of $B^2$. Hence the number of quadruples that correspond to incidences occurring on those curves $\gamma_{a,b}$ is $O(|A||B|)$, and we trim $Q_0$ further, by removing from it all these quadruples. This yields the identity
\begin{equation}\label{eq:idg1}
f(a,x)-f(b,x)\equiv-\lambda g(x,x).
\end{equation}
Note that, by the definition of $\ell$ and by our assumption that $\tilde c_\ell(a)\neq\tilde c_\ell(b)$, the polynomial on the left-hand side of (\ref{eq:idg1}) is of degree exactly $\ell$. Let $g_1(x):=g(x,x)$. Comparing the degrees of the polynomial on both sides of (\ref{eq:idg1}), we get $\ell= \deg g_1\le\deg g=d_v-e<d_v$, where $e$ is the degree of $c_k$. In particular, the leading term of $f(a,x)-f(b,y)$, which is of degree $d_v$, is independent of $a$ and $b$. 

By (\ref{eq:idxy}), we have
$$
[f(a,x)-f(b,y)]^* \equiv [c_k(x)-c_k(y)]^*[g(x,y)]^*,
$$
or
\begin{equation}\label{eq:ltg}
[g(x,y)]^*\equiv \frac{[f(a,x)-f(b,y)]^* }{[c_k(x)-c_k(y)]^*}.
\end{equation}
As just argued, the numerator (and certainly also the denominator) of the right-hand side of (\ref{eq:ltg}) is independent of $a$ and $b$. Hence, up to a (non-zero) multiplicative constant, we have 
\begin{equation}\label{eq:ltg2}
[g(x,y)]^*\equiv\frac{x^{d_v}-y^{d_v}}{x^e-y^e}
\equiv x^{d_v-e}+x^{d_v-2e}y^e+\cdots+x^{e}y^{d_v-2e}+y^{d_v-e}.
\end{equation}
In particular, substituting $y=x$ again, $[g(x,x)]^*$ becomes $\frac{d_v}ex^{d_v-e}\neq 0$. 
Comparing the leading terms on both sides of (\ref{eq:idg1}) yields
$$
[f(a,x)-f(b,x)]^* \equiv -\lambda[g(x,x)]^*,
$$
or
$$
\tilde c_\ell(a)-\tilde c_\ell(b)= -\lambda d_v/e.
$$
So, for every $(x,y)\in\gamma'$, and every $(a,b)\in S(\gamma')$, we have
$$
\tilde c_\ell(a)-\tilde c_\ell(b)= -(d_v/e)(c_k(x)-c_k(y)),
$$
or
\begin{equation}\label{eq:hsum}
\tilde c_\ell(a)+(d_v/e) c_k(x)= \tilde c_\ell(b)+(d_v/e)c_k(y).
\end{equation}
Put
$$
h(u,v):=\tilde c_\ell(u)+(d_v/e) c_k(v),
$$
and observe that $h$ does not depend on $\gamma'$, that $\deg h\le d\le 2d^2$, and that $h$ satisfies the property assumed in the proposition. 

%----------------------Case: k=d_u----------------------
\paragraph{The case $k=d_u$.}

First note that, arguing as in the previous case, and by the symmetry of the two setups, $\ell<d_v$ implies $k<d_u$. Hence, we can assume that in this case we also have $\ell=d_v$.
We replace $(\xi_3,\eta_3)$ in~(\ref{eqco}) by a generic point $(\xi,\eta)$ along $\gamma'$, and conclude that all points $(\xi,\eta)$ on $\gamma'$, except possibly for some finite discrete subset (recall the previous comment concerning isolated points), satisfy
\begin{equation} \label{la-primal}
c_{d_u}(\eta) = \mu c_{d_u}(\xi) ,
\end{equation}
for some fixed parameter $\mu$, which depends on $\gamma'$. (In contrast, $c_{d_u}$ itself depends only on $f$ and not on $\gamma'$.) 

Consider now the system
\begin{align*}
f(a,x) - f(b,y) & = 0 \\
\mu c_{d_u}(x)- c_{d_u}(y) & = 0 , 
\end{align*}
of polynomials in $x,y$, which has infinitely many solutions for each pair $(a,b)\in S(\gamma')$ (they both vanish on $\gamma'$). That is, the polynomials $f(a,x) - f(b,y)$ and $\mu c_{d_u}(x)- c_{d_u}(y)$ have a common factor. The corresponding system of leading terms is (after dividing the second equation by the leading coefficient of $c_{d_u}$)
\begin{align*}
& \tilde c_{d_v}(a) x^{d_v} - \tilde c_{d_v}(b)y^{d_v} \\
& \mu x^{e'} - y^{e'} , 
\end{align*}
for a suitable exponent $e'\ge 1$. By discarding at most $O(|A||B|)$ further quadruples from $Q_0$, we may assume that neither of the coefficients $\tilde c_{d_v}(a)$, $\tilde c_{d_v}(b)$ is zero. Indeed, the equation $\tilde c_{d_v}(x)=0$ has at most $d_u$ solutions in $\R$, and hence there are at most $2d_u|A|$ pairs $(a,b)$ in $A^2$ such that one of $a, b$ is one of these solutions. For each pair $(a,b)$ of this form, by the Schwartz-Zippel lemma (Lemma~\ref{SZ}), $\gamma_{a,b}$ is incident to at most $O(|B|)$ points of $B^2$. Hence the total number of quadruples involving those curves is at most $O(|A||B|)$, and we remove all of them from $Q_0$.
Then Lemma~\ref{common} implies that
$$
\left( \frac{ \tilde c_{d_v}(a)  }{ \tilde c_{d_v}(b) } \right)^{e'} = \mu^{d_v} =
\left( \frac{ c_{d_u}(y) }{ c_{d_u}(x) } \right)^{d_v} .
$$
That is, 
\begin{equation}\label{eq:hmul}
\tilde c_{d_v}(a)^{e'} c_{d_u}(x)^{d_v} = 
\tilde c_{d_v}(b)^{e'} c_{d_u}(y)^{d_v} , 
\end{equation}
for each $(x,y)\in\gamma'$ and each $(a,b)\in S(\gamma')$. 
Put
$$
h(u,v):=\tilde c_{d_v}(u)^{e'} c_{d_u}(v)^{d_v}
$$
and observe that $h$ does not depend on $\gamma'$, that $\deg h\le 2d^2$, and that $h$ satisfies the property assumed in the proposition.

In either of the two cases, substituting $(x,y)=(p,q)$ in (\ref{eq:hsum}) or in (\ref{eq:hmul}) yields $h(a,p)=h(b,q)$, as asserted.
$\Box$

%-------------------------------------------------------------------------Applications---------------------------------------------------------------
\section{Applications}\label{sec:appl}
\subsection{Directions determined by a planar point set}
For a finite point set $P\subset\R^2$ we denote by $S(P)$ the number of distinct directions determined by pairs of points of $P$. The study of sets that determine few distinct directions was initiated by Scott \cite{Sco}. He conjectured that $S(P) \ge|P|-1$ for any non-collinear planar point set. This was settled in the affirmative by Ungar~\cite{Ung}. Sets for which equality holds are called {\em critical} by Jamison \cite{Jam} and those with one additional direction, i.e., sets satisfying $S(P)=|P|$, are called {\em near-critical}. Jamison gives an overview of the known critical and near-critical configurations in the Euclidean plane and, among other results, characterizes critical or near-critical configurations that lie in the union of two or three straight lines.

Little is known about plane sets with $S(P)=|P|+1$, let alone $S(P)\le c|P|$, for some constant $c>0$. One result in this direction, due to Elekes~\cite{El3}, is that the Jamison configurations are essentially the only structures that can satisfy even the much weaker requirement $S(P)\le c|P|$, provided that $P$ contains $\alpha|P|$ collinear points, where $\alpha$ is a sufficiently large fraction (see \cite{El3} for the exact statement). In the same paper Elekes also proves the following theorem.
\begin{thm}[Elekes~\cite{El3}]
Let $\gamma\subset\R^2$ be a curve of the form $y = f(x)$, where $f$ is some constant-degree polynomial, and $\deg f\ge 3$. Then, for any finite point set $P\subset \gamma$, we have $S(P)=\omega(|P|)$.
\end{thm}

Theorem \ref{main2} (more precisely, Corollary \ref{cor:main}) yields the following significant sharpening of this result.
\begin{thm}
Let $\gamma\subset\R^2$ be a curve of the form $y = f(x)$, where $f$ is a constant-degree polynomial, and $\deg f\ge 3$. Then, for any finite point set $P\subset \gamma$, we have $S(P)=\Omega(|P|^{4/3})$.
\end{thm}
\noindent{\bf Proof.}
Consider the polynomial function $g(x,y):=\frac{f(x)-f(y)}{x-y}$. It is shown in \cite{El3} that $g$ is not of one of the special forms in Theorem~\ref{main2}. The asserted bound then follows from Corollary \ref{cor:main}.
$\Box$ 

For completeness, we mention the following recent result of Elekes and Szab\'o~\cite{ESz2}.
\begin{thm}[Elekes and Szab\'o~\cite{ESz2}]\label{thm:ESz2}
Let $\gamma\subset\R^2$ be an irreducible constant-degree algebraic curve. Then, either $\gamma$ is a conic section, or, for any finite point set $P\subset \gamma$, we have $S(P)=\omega(|P|)$.
\end{thm}

It is easy to construct examples of a finite set $P$ that lies on a conic section, such as a circle or a pair of lines, that determines only $\Theta(|P|)$ distinct directions.

For the proof, Elekes and Szab\'o use their earlier result from~\cite{ESz}, also mentioned in the introduction, which deals with implicit surfaces of the form $F(x, y, z) = 0$, where $F$ is some constant-degree trivariate polynomial. Extending Theorem~\ref{main} to such surfaces, which we believe is possible (see concluding section for more details), will sharpen the `gap' given by Theorem~\ref{thm:ESz2}.

\subsection{Distinct distances: Special configurations}
For a finite point set $P$ (lying in some Euclidean space), we denote by $D(P)$ the number of distinct distances determined by pairs of points of $P$. In~\cite{Cha}, Charalambides proved the following theorem.
\begin{thm}[Charalambides~\cite{Cha}]\label{thm:Cha}
Let $\gamma\subset\R^d$ be a constant-degree irreducible algebraic curve, which is not an algebraic helix. Then, for any finite point set $P\subset\gamma$, $D(P)=\Omega(|P|^{5/4})$.
\end{thm}
\noindent (Here an {\it algebraic helix} is either a line, or a curve that, up to a rigid motion, admits a parameterization of the form $(u_1,\ldots,u_k)\mapsto (\alpha_1\cos u_1,\alpha_1\sin u_1,\ldots,\alpha_k\cos u_k,\alpha_k\sin u_k)\in\R^{2k}\subset\R^d$, for some parameter $k\le d/2$.)

For the special case $d=2$, Pach and de Zeeuw \cite{PdZ} managed to improve the lower bound obtained in Theorem~\ref{thm:Cha} as follows; their result generalizes the bound (\ref{eq:dist}) mentioned in the introduction.
\begin{thm}[Pach and de Zeeuw~\cite{PdZ}]\label{thm:PdZ}
Let $\gamma\subset\R^2$ be a constant-degree irreducible algebraic curve which is not a line or a circle. Then, for any finite point set $P\subset\gamma$, $D(P)=\Omega(|P|^{4/3})$.
\end{thm}

Using our machinery, we obtain the same lower bound of $\Omega(|P|^{4/3})$ for points on a curve in an arbitrary (constant) dimension $d$, improving the bound given in Theorem~\ref{thm:Cha}; our result however is somewhat restricted, because it requires the curve $\gamma$ to have a polynomial parameterization.
\begin{thm}~\label{thm:dist}
Let $\gamma\subset\R^d$ be a curve of the form $\gamma(t)=(x_1(t),\ldots,x_d(t))$, for $t\in\R$, where $x_1(t),\ldots,x_d(t)$ are some constant-degree polynomials. Then, either $\gamma$ is a line, or, for any finite point set $P\subset\gamma$, $D(P)=\Omega(|P|^{4/3})$.
\end{thm}
\noindent{\bf Proof.}
Consider the bivariate polynomial function 
$$
f(t,s):=\|\gamma(t)-\gamma(s)\|^2=\sum_{i=1}^d(x_i(t)-x_i(s))^2,\quad \text{for}~t,s\in\R.
$$ 
By shifting the coordinate frame, we may assume that $x_i(0)=0$, and we also assume that $x_i(t)$ is not identically zero, for each $i=1,\ldots,d$; coordinates for which $x_i\equiv 0$ do not affect the function $f$ and can simply be ignored. Suppose that $f$ is of one of the forms $h(\varphi(t)+\psi(s))$, or $h(\varphi(t)\cdot\psi(s))$, for some univariate polynomials $h,\varphi,\psi$. We claim that in this case $\gamma$ must be a line (the converse statement is easy to verify).

Consider first the multiplicative special form. That is, assume that
$$
\sum_{i=1}^d(x_i(t)-x_i(s))^2=h(\varphi(t)\psi(s)),
$$
for suitable univariate polynomials $h,\varphi,\psi$. Substituting $t=s$, the left-hand side in the above identity is zero, and hence we must have that $\varphi(t)\psi(t)\equiv x_0$, where $x_0$ is a real root of $h$. However, this can occur only if the polynomials $\varphi,\psi$ are both constants, that is, only if $h(\varphi(t)\psi(s))$ is a constant independent of $t$ and $s$. Since this quantity corresponds to the distance between two points, represented by the parameters $t,s$, on the (one-dimensional) curve $\gamma$, this yields a contradiction.

Next consider the additive special form. That is, assume that
$$
\sum_{i=1}^d(x_i(t)-x_i(s))^2=h(\varphi(t)+\psi(s)),
$$
for $h,\varphi,\psi$, as above. Substituting $t=s$, as above, we must have that $\varphi(t)+\psi(t)\equiv x_0$, where $x_0$ is a real root of $h$. The last identity then becomes
\begin{equation}\label{eq:hx0}
\sum_{i=1}^d(x_i(t)-x_i(s))^2=h(\varphi(t)-\varphi(s)+x_0).
\end{equation}
Moreover, the multiplicity of $x_0$ (as a root of $h$) is at least two (because this is the multiplicity of the factor $t-s$ of the polynomial on the left-hand side of this identity).

Taking derivatives on both sides of (\ref{eq:hx0}) twice, once with respect to $t$ and then with respect to $s$, yields
\begin{equation}
2\sum_{i=1}^dx_i'(t)x_i'(s)=h''(\varphi(t)-\varphi(s)+x_0)\varphi'(t)\varphi'(s).
\end{equation}
Comparing the leading terms, we see that $h''$ must be a constant. Indeed, the leading term of the left-hand side, divided by $\varphi'(t)\varphi'(s)$, is of the form $\alpha t^es^e$, for some integer $e\ge 0$, and some constant $\alpha$. Hence the leading term of $h''(\varphi(t)-\varphi(s)+x_0)$ must also have this form. Let $e'$ denote the degree of $h$ (as a univariate polynomial). The leading term of $h''(\varphi(t)-\varphi(s)+x_0)$ (as a bivariate polynomial of $t$ and $s$) is, up to a constant multiplicative factor, the same as the leading term of $(\varphi(t)-\varphi(s)+x_0)^{e'}$. Then clearly, in order to have the form $\alpha t^es^e$, it must be that $e'=0$. Hence $h''$ is a constant, and thus $h$ is a polynomial of degree two. Since $x_0$ is a multiple root of $h$, this implies
\begin{equation}\label{eq:com}
\sum_{i=1}^d(x_i(t)-x_i(s))^2=(\varphi(t)-\varphi(s))^2.
\end{equation}

We have assumed that neither of the polynomials $x_i$ has a constant term, and we may assume this also holds for $\varphi$. We then get 
$$
\sum_{i=1}^dx_i^2(t)+\sum_{i=1}^dx_i^2(s)-2\sum_{i=1}^dx_i(t)x_i(s)\equiv \varphi^2(t)+\varphi^2(s)-2\varphi(t)\varphi(s).
$$
This in turn implies that $\sum_{i=1}^dx_i^2(t)\equiv \varphi^2(t)$ (all the other terms are divisible by $s$), and thus $\sum_{i=1}^dx_i(t)x_i(s)\equiv \varphi(t)\varphi(s)$.
That is, the scalar product of the two vectors $(x_1(t),\ldots,x_d(t))$ and $(x_1(s),\ldots,x_d(s))$ is equal to the product of their lengths, so these vectors must be parallel, for every pair $s,t\in\R$. In other words, $\gamma$ must be a line through the origin. Removing our assumption that $x_i(0)=0$, $\gamma$ can be any line, as claimed.

Hence, if $\gamma$ is not a line, $f$ cannot have one of the special forms, and Corollary~\ref{cor:main} implies that $D(P)=\Omega(|P|^{4/3})$, as asserted.
$\Box$

Recently, Bronner et al.~\cite{BSS} considered a bipartite version of the distinct distances problem, where we are given two finite point sets $P_1,P_2$ in $\R^d$, with $d\ge 3$, and the points of $P_1$ are contained in a line $\ell$ (and without any restriction on the points of $P_2$). They showed that the number of distinct distances spanned by pairs of points from $P_1\times P_2$ is 
$$
\Omega\left(\min\{|P_1|^{4/5}|P_2|^{2/5},|P_1|^2,|P_2|^2\}\right),
$$
unless many of the points of $P_2$ lie either on a cylinder, with $\ell$ as its axis, or on an hyperplane orthogonal to $\ell$ (see \cite{BSS} for the exact statement, and for more results of this kind). 

\subsection{Sum-product-type estimates}
Variants of the sum-product problem have been studied intensively since the work of Erd\H{o}s and Szemer\'edi~\cite{ErSz}, where it was shown that there exists $c> 0$ such that for any finite set $A\subset\Z$, one has 
$$
|A + A|+|A \cdot  A|\ge |A|^{1+c},
$$
where $A+A=\{u+v\mid u,v\in A\}$, and $A\cdot A=\{uv\mid u,v\in A\}$. Much of the subsequent extensive work aimed either to give an explicit (lower) bound for $c$, or to derive generalizations of the sum-product lower bound. The currently best known lower bound is due to Solymosi~\cite{Soly}, and asserts that, for any finite set $A\subset\R$, one has
$$
|A + A|+|A \cdot  A|\ge \frac{|A|^{4/3}}{2\lceil\log|A|\rceil^{1/3}}.
$$

One of the significant generalizations of this problem is the work by Elekes et al.~\cite{ENR} who showed that, for any given finite set $A\subset \R$, and a strictly convex (or concave) function $f$ defined on an interval
containing $A$, one has
\begin{equation}\label{eq:sumconv}
|A + A|+|f(A) + f(A)|=\Omega(|A|^{5/4}).
\end{equation}

%Elekes et al. actually proved the following more general result.
%\begin{thm}[Elekes et al.~\cite{ENR}]\label{ENR}
%Let $A\subset\R$ be a finite set, with $|A|=N$, and let $f:\R\to\R$ be a strictly convex (or
%concave) function, defined on an interval containing $A$. Then, for any pair of sets $C,D\subset\R$ satisfying $|C||D|\ge N$, we have
%\begin{equation}\label{eq:ENR}
%|A+C||f(A)+D|=\Omega(N^{3/2}(|C||D|)^{1/2}).
%\end{equation}
%\end{thm}
%Substituting $C=A$, $D=f(A)$ in (\ref{eq:ENR}) gives (\ref{eq:sumconv}) (here the strict convexity or concavity of $f$ implies that $|D|=\Omega(|A|)$).

The bound (\ref{eq:sumconv}) was recently improved by Li and Roche-Newton~\cite{LRN}; their result is based on a breakthrough work by Schoen and Shkredov~\cite{ScSh}. 
\begin{thm}[Li and Roche-Newton~\cite{LRN}]\label{thm:LR}
Let $f$ be a continuous strictly convex or concave function on the reals. Let $A,C\subset\R$ be finite sets, such that $|A|=|C|=N$. Then
\begin{equation}\label{eq:LR}
|A+A|+|f(A)+C|=\Omega\left(\frac{N^{24/19}}{\log^{2/19}N}\right).
\end{equation}
\end{thm}

Theorem~\ref{thm:LR} immediately implies the following lemma (a similar argument was used in Green and Tao~\cite{GT}). 
\begin{lem}\label{lem:LR}
Let $A,C\subset\R$ be finite sets, such that $|A|=|C|=N$, and let $f:\R\to\R$ be a continuous function. Suppose that there exist $x_1<x_2<\cdots<x_c$, for some constant index $c\ge2$, such that $f$ is strictly concave or convex on each open interval $(x_i,x_{i+1})$. Then
\begin{equation}
|A+A|+|f(A)+C|=\Omega\left(\frac{N^{24/19}}{\log^{2/19}N}\right).
\end{equation}
\end{lem}
\noindent{\bf Proof.}
Let $A'$ denote the largest set among the sets $A_i=A\cap(x_i,x_{i+1})$, which is of cardinality at least $N/c$, and similarly let $C'$ denote the largest set among the sets $C_i=C\cap(x_i,x_{i+1})$, which is also of cardinality at least $N/c$. Applying Theorem~\ref{thm:LR} to $A'$ and $C'$ yields
$$
|A+A|+ |f(A)+C|\ge |A'+A'|+ |f(A')+C'|=\Omega\left(\frac{N^{24/19}}{\log^{2/19}N}\right),
$$
as asserted.
$\Box$

Recently, Shen~\cite{She} proved the following generalization of (\ref{eq:sumconv}).
\begin{thm}[Shen~\cite{She}]\label{shen}
Let $f$ be a bivariate constant-degree real polynomial. Then either $f$ is of the form $f(x,y)=h(ax+by)$, for some univariate polynomial $h$ and constants $a,b\in\R$, or, for any finite set $A\subset\R$, one has
$$
|A+A|+|f(A,A)| =\Omega(|A|^{5/4}).
$$
\end{thm}
In view of Corollary~\ref{cor:main},  Shen's result is interesting only in the case where $f$ is of one of the special forms from Theorem~\ref{main2}, since in the complementary case we always have $|f(A,A)|=\Omega(|A|^{4/3})$. Moreover, for functions $f$ having one of the special forms, an improved bound for Theorem~\ref{shen} follows from Theorem~\ref{thm:LR} (and Lemma~\ref{lem:LR}), as is next shown. Since the overall conclusion is somewhat asymmetric, let us state it explicitly. 
\begin{cor}\label{cor:impshen}
Let $f$ be a bivariate constant-degree real polynomial. If $f$ is not of one of the forms $f(x,y)=h(\varphi(x)+\psi(y))$, or $f(x,y)=h(\varphi(x)\cdot\psi(y))$, for some univariate polynomials $h,\varphi,\psi$, then, for any finite set $A\subset\R$, one has
$$
|A+A|+|f(A,A)| =\Omega(|A|^{4/3}).
$$
Otherwise, either $f$ is of the form $f(x,y)=h(ax+by)$, for some constants $a,b\in\R$, or, for any finite set $A\subset\R$, one has
$$
|A+A|+|f(A,A)| =\Omega\left(\frac{|A|^{24/19}}{\log^{2/19}|A|}\right).
$$
\end{cor}

%Actually, our analysis implies the following more general result, involving two different sets (of potentially different cardinalities), which does not appear in~\cite{She}. 
%\begin{cor}
%Let $f$ be a bivariate constant-degree real polynomial. If $f$ is not of one of the forms $f(x,y)=h(\varphi(x)+\psi(y))$, or $f(x,y)=h(\varphi(x)\cdot\psi(y))$, for some univariate polynomials $h,\varphi,\psi$, then, for any pair of finite sets $A,B\subset\R$, one has
%$$
%|A+B|+|f(A,B)| =\Omega\left(|A|^{2/3}|B|^{2/3}+|A|+|B|\right).
%$$
%Otherwise, either $f$ is of the form $f(x,y)=h(ax+by)$, for some univariate polynomial $h$ and constants $a,b\in\R$, or, for any two finite sets $A,B\subset\R$, with $|A|^{1/2}\le |B|\le|A|^2 $, one has
%$$
%|A+B|+|f(A,B)| =\Omega\left(\min\left\{\;|A|^{3/4}|B|^{1/2}\;,~\;|A|^{1/2}|B|^{3/4}\;\right\}\right).
%$$
%\end{cor}
\noindent{\bf Proof.}
If $f$ is not of one of the forms $f(x,y)=h(\varphi(x)+\psi(y))$, or $f(x,y)=h(\varphi(x)\cdot\psi(y))$, Corollary~\ref{cor:main} implies 
$$
|A+A|+|f(A,A)|\ge |f(A,A)| =\Omega\left(|A|^{4/3}\right).
$$
To treat the complementary case, assume that $f$ has one of the above forms, and that $\varphi$, say, is a nonlinear polynomial (the special form of $f$ asserted in Theorem~\ref{shen} is merely an equivalent way of saying that $h$ has the additive form and that both $\varphi$ and $\psi$ are linear). We may assume that $\psi$ is not a constant, for otherwise $f$ is independent of $y$ and thus can be written as $h(ax+by)$, with $a=1$, $b=0$. Clearly, it also suffices to assume that $h$ is non-constant. If $f$ has the form $f(x,y)=h(\varphi(x)+\psi(y))$, we apply Lemma \ref{lem:LR} to the function  $\varphi$, and to the sets $A$ and $C:=\psi(A)$; since $\varphi$ is a nonlinear constant-degree polynomial, and $\psi$ is a constant-degree polynomial (which is non-constant), it is easy to see that the conditions of the lemma are satisfied. If $f$ is of the multiplicative special form $f(x,y)=h(\varphi(x)\psi(y))$, we again apply Lemma \ref{lem:LR}, this time to  the function $\ln${\small$\circ$}$\varphi$ and the sets $A$ and $C:=\ln(\psi(A))$. In the former case we obtain 
$$
|A+A|+|\varphi(A)+\psi(A)|=\Omega\left(\frac{N^{24/19}}{\log^{2/19}N}\right).
$$
Since $f(A,A)=h(\varphi(A)+\psi(A))$ and $h$ is of constant degree, we have $|f(A,A)|=\Omega(|\varphi(A)+\psi(A)|)$, and the asserted bound follows. In the latter case a similar argument shows that 
$$
|A+A|+|f(A,A)|=\Omega(|A+A|+|\varphi(A)\cdot\psi(A)|)=\Omega\left(\frac{N^{24/19}}{\log^{2/19}N}\right),
$$
as asserted.
$\Box$

The following theorem is also taken from~\cite{LRN}; it considers $A-A$ instead of $A+A$, and provides a sharper lower bound.
\begin{thm}[Li and Roche-Newton~\cite{LRN}]\label{thm:LR2}
Let $f$ be a continuous strictly convex or concave function on the reals. Let $A,C\subset\R$ be finite sets, such that $|A|=|C|=N$. Then
\begin{equation}\label{eq:LR}
|A-A|+|f(A)+C|=\Omega\left(\frac{N^{14/11}}{\log^{2/11}N}\right).
\end{equation}
\end{thm}

This allows us to prove the following variant of Shen's theorem, with a better lower bound (larger than our improved bound obtained in Corollary~\ref{cor:impshen}). The proof is similar to the one of Corollary~\ref{cor:impshen}.
\begin{cor}
Let $f$ be a bivariate constant-degree real polynomial. If $f$ is not of one of the forms $f(x,y)=h(\varphi(x)+\psi(y))$, or $f(x,y)=h(\varphi(x)\cdot\psi(y))$, for some univariate polynomials $h,\varphi,\psi$, then, for any finite set $A\subset\R$, one has
$$
|A- A|+|f(A,A)| =\Omega(|A|^{4/3}).
$$
Otherwise, either $f$ is of the form $f(x,y)=h(ax+by)$, for some constants $a,b\in\R$, or, for any finite set $A\subset\R$, one has
$$
|A-A|+|f(A,A)| =\Omega\left(\frac{|A|^{14/11}}{\log^{2/11}|A|}\right).
$$
\end{cor}
\noindent{\it Remark.} In the special case where $f(x,y) = h(ax + by)$, with $h$, $a$, and $b$ as above, one can construct sets $A$ of arbitrarily large size so that $|A\pm A| + |f(A,A)|$ is $\Theta(|A|)$, provided that $b/a$ satisfies certain properties, such as being rational. Results regarding this issue, for the special case where $A \subset \Z$, can be found in \cite{Buk, CHS} (see also~\cite{Pla} for some results of this kind for finite fields).

%----------------------------------------------------------------------Conclusion-------------------------------------------------------------------------------------------------------------------
\section{Conclusion}\label{sec:con}
At a high level, there are some common features of the analysis
of Elekes and R\'onyai~\cite{ER00} and ours, but there are considerable differences
in the actual analysis (and results). At the risk of making the comparison
somewhat informal and imprecise, we list a few similarities and differences.

\medskip
\noindent (i) Both techniques double count ``quadruples'', or rather
``quintuples'' of various kinds. For example, in our second proof of 
Theorem~\ref{main2} we consider quintuples 
$(a,b,p,q,c)\in A^2\times B^2\times C$ such that $f(a,p)=f(b,q)=c$ (the 
parameter $c$ is implicit in our setup). In contrast, Elekes and R\'onyai 
consider quintuples of the form $(a,b_1,b_2,c_1,c_2)\in A\times B^2\times C^2$,
such that $f(a,b_1)=c_1$ and $f(a,b_2)=c_2$.

\medskip
\noindent (ii) In both cases the quintuples are interpreted as incidences between
points and curves in a suitable parametric plane. The reductions are different, though,
and the remainders of the proofs are a consequence of the parameterization. Elekes and
R\'onyai obtain curves of the form $\{(f(t,b_i),f(t,b_j)) \mid b_i,b_j\in B\}$, which
are rationally or polynomially parameterizable. Our curves are different (and the
curves appearing in the two proofs of the theorem are also different from one another).

\medskip
\noindent (iii) One notable difference is that Elekes and R\'onyai's goal was only
to establish a dichotomy between the case where (in our notation) $M$ is quadratic and
$f$ has one of the special forms, and the complementary case. They did not set up to
obtain a concrete `gap' between the two cases, as we do in this paper. (Such a (weaker)
gap has been obtained later, by Elekes and Szab\'o~\cite{ESz}, in their treatment of a 
more general setup.)

\medskip
\noindent (iv) Both proofs use rather elementary algebra of polynomials, of different sorts.

A recent study of Tao~\cite{Tao} derives similar results for bivariate polynomials over finite fields. The methodology in his analysis resembles ours (and the one in \cite{ER00}), in the sense of counting quadruples (albeit of a somewhat different sort).

An obvious direction for further research is to extend the machinery developed in this paper to the more general setup of Elekes and Szab\'o~\cite{ESz}, involving the vanishing of a trivariate polynomial $F(x,y,z)$ on a three-dimensional grid $A\times B\times C$. The general high-level approach is clear: We can consider the set $Q$ of quadruples $(a,a',b,b')\in A^2\times B^2$, such that there exists $c\in C$ satisfying $F(a,b,c)=F(a',b',c)=0$ (compared with what has just been noted, this actually counts the quintuples $(a,b,a',b',c)$), relate the number $|Q|$ of such quadruples, via the Cauchy-Schwarz inequality, to $M$, the number of zeros of $F$ on the grid, and then interpret each quadruple as an incidence between, e.g., the point $(b,b')$ and a suitable curve $\gamma_{a,a'}$, defined in analogy with the curves of Section~\ref{sec:part2}. Again, the main technical hurdle is to handle situations where too many of these curves (or of their duals, flipping the roles of $A$ and $B$) overlap in a common irreducible component. That is, the challenge is to show that if this is not the case then $Q$ can be bounded via a standard incidence bound, as we did above, and then the bound $M=O(n^{11/6})$ (or the more elaborate bound of Theorem~\ref{main2}) would follow, and if there exist overlaps of large multiplicity, then $F$ must be special, e.g., in the sense of \cite{ESz}. 

We believe that our analysis can also be applied over the complex field, and leave this extension as (what we hope would be an easy) open problem. Most of the analysis carries over to the complex setting with hardly any change, except for certain issues which require a more careful adaptation. One such issue is the use of the planar incidence technique of Sz\'ekely~\cite{Sz97}. In the complex case this would have to be replaced by a different technique, similar to the recent proofs of the complex Szemer\'edi-Trotter theorem due to Solymosi and Tao~\cite{SoT} and to Zahl~\cite{Zah} (see also T\'oth~\cite{Tot}).

Another interesting challenge is to extend the result to higher-dimensional grids; see Schwartz et al.~\cite{SSdZ} for an initial attempt in this direction for four-dimensional grids. An even more challenging direction would be to extend the analysis to cases where the constituent sets $A$, $B$, $C$ of the grid are not one-dimensional. In these cases the problem would translate to incidences between points and higher-dimensional varieties, typically, points and two-dimensional varieties in $\reals^4$ (when $A$ and $B$ are sets of points in the plane).

Another interesting project is to obtain a sharp calibration of the dependence of the bounds in this paper on the degree of $f(x,y)$. For example, our results and those of~\cite{PdZ}, show that the number of distinct distances between $n$ points on a constant-degree curve (which is neither a line or a circle) in the plane is $\Omega(n^{4/3})$. On the other hand, for any set of $n$ points in the plane there exists a curve of degree $d=O(\sqrt{n})$ that passes through all the points (e.g., see~\cite{GK1}), and then the nearly linear upper bound on the number of distinct distances in the grid construction of Erd{\H o}s~\cite{Er46} suggests that we will not
be able to prove a superlinear lower bound when $d=\Theta(\sqrt{n})$. Is there any hope in deriving a lower bound that depends on $d$, and interpolate between the two extreme situations noted above?

Another open problem is to improve the bound on $M$ in Theorems \ref{main} and \ref{main2}. We are not aware of any non-trivial lower bound for $M$, and suspect it to be much smaller.

Finally, it would be interesting to find additional applications of the results of this paper.

%------------------------------------------------------------------Bibliography---------------------------------------------------------------------------------------------------------


\begin{thebibliography}{}

\bibitem{Aya}
M. Ayad,
Sur les polyn\^omes $f(X,Y)$ tels que $K[f]$ est int\'egralement ferm\'e dans $K[X,Y]$,
\emph{Acta Arithmetica} 105 (2002), 9--28.

\bibitem{BSS}
A. Bronner, M. Sharir and A. Sheffer,
Distinct distances on a line and a curve,
manuscript, 2014.

\bibitem{Buk}
B. Bukh,
Sums of dilates,
{\em Combin. Probab. Comput.}, 17(5) (2008), 627--639. 

\bibitem{Cha}
M. Charalambides,
Exponent gaps on curves via rigidity,
in arXiv:1307.0870 (2013).

\bibitem{CHS}
J. Cilleruelo, Y. Hamidoune and O. Serra,
 On sums of dilates,
{\em Combin. Probab. Comput.}, 18(6) (2009), 871--880.

\bibitem{CLO}
D. A. Cox, J. Little and D. O'Shea,
{\it Using Algebraic Geometry},
Springer-Verlag, 2nd Edition, Heidelberg 2005.

\bibitem{El}
G. Elekes, 
Sums versus products in number theory, algebra and Erd{\H o}s geometry--A survey,
in {\it Paul Erd{\H o}s and his Mathematics} II,
Bolyai Math. Soc., Stud. 11, Budapest, 2002, pp.~241--290.

\bibitem{El3}
G. Elekes, 
On linear Combinatorics III,
{\em Combinatorica}, 19(1) (1999), 43--53.

\bibitem{ENR}
G. Elekes, M. Nathanson and I. Ruzsa, 
Convexity and sumsets, 
\emph{J. Number Theory} 83(2) (2000), 194--201.

\bibitem{ER00}
G.\ Elekes and L.\ R\'onyai, 
A combinatorial problem on polynomials and rational functions,
{\it J.\ Combin.\ Theory Ser.\ A} 89 (2000), 1--20.

\bibitem{ESSz}
G. Elekes, M. Simonovits and E. Szab\'o,
A combinatorial distinction between unit circles and straight lines:
How many coincidences can they have?
{\it Combin. Probab. Comput.} 18 (2009), 691--705.

\bibitem{ESz}
G. Elekes and E. Szab\'o,
How to find groups? (and how to use them in Erd{\H o}s geometry?),
{\it Combinatorica} 32 (2012), 537--571.

\bibitem{ESz2}
G. Elekes and E. Szab\'o,
On triple lines and cubic curves: The Orchard Problem revisited,
in arXiv:1302.5777 (2013).

\bibitem{Er46}
P.~Erd{\H o}s, 
On a set of distances of $n$ points, 
{\it Amer. Math.  Monthly} 53 (1946), 248--250.

\bibitem{ErSz}
P. Erd\H{o}s and E. Szemer\'edi,
On sums and products of integers,
in: P. Erd\H{o}s, L. Alp\'ar, G. Hal\'asz, and A. S\'ark\"ozy, editors, {\it Studies in Pure Mathematics, To the Memory of Paul
Tur\'an}, 1983, Birkh\"auser Verlag, Basel, pp. 213--218.

\bibitem{GT}
B. Green and T. Tao,
On sets defining few ordinary lines,
in arXiv:1208.4714 (2012).

\bibitem{GK1}
L. Guth and N. H. Katz,
Algebraic methods in discrete analogs of the Kakeya problem,
{\it Advances Math.} 225 (2010), 2828--2839.
Also in arXiv:0812.1043v1.

\bibitem{Jam}
R. E. Jamison,
Planar configurations which determine few slopes,
{\em Geometriae Dedicata} 16 (1984), 17--34.

\bibitem{LRN}
L. P. Li and O. Roche-Newton,
Convexity and a sum-product type estimate,
{\em Acta Arithmetica} 156.3 (2012), 247--255.


\bibitem{Mil64}
J. Milnor, 
On the Betti numbers of real varieties,
{\it Proc. Amer. Math. Soc.}, 
15(2) (1964), 275--280.

\bibitem{PA95}
J. Pach and P. K. Agarwal,
{\it Combinatorial Geometry},
Wiley-Interscience, New York 1995.

\bibitem{PS98}
J. Pach and M. Sharir,
On the number of incidences between points and curves,
{\it Combin. Probab. Comput.} 7 (1998), 121--127.

\bibitem{PdZ}
J. Pach and F. de Zeeuw,
Distinct distances on algebraic curves in the plane,
in arXiv:1308.0177 (2013).

\bibitem{Pla}
A. Plagne, 
Sums of dilates in groups of prime order,
{\it Combin. Probab. Comput.}, 20(6) (2011), 867--873.

\bibitem{RSS}
O. E. Raz, M. Sharir, and J. Solymosi,
On triple intersections of three families of unit circles,
manuscript, 2013.

\bibitem{ScSh}
T. Schoen and I. Shkredov,
On sumsets of convex sets,
{\em Combin. Probab. Comput.}, 20(6) (2011), 793--798.

\bibitem{Sch80}
J. Schwartz,
Fast probabilistic algorithms for verification of polynomial identities,
{\it J. ACM} 27(4) (1980), 701--717.

\bibitem{SSdZ}
R. Schwartz, J. Solymosi, and F. de Zeeuw,
Extensions of a result of Elekes and R\'onyai,
{\it J. Combin. Theory Ser. A} 120(7) (2013), 1695--1713.

\bibitem{Sco}
P. R. Scott.
On the sets of directions determined by $n$ points,
{\em Amer. Math. Monthly} 77 (1970), 502--505.


\bibitem{SSS}
M. Sharir, A. Sheffer, and J. Solymosi,
Distinct distances on two lines,
{\it J. Combin. Theory Ser. A} 20 (2013), 1732--1736.

\bibitem{SS}
M. Sharir and J. Solymosi,
Distinct distances from three points, 
in arXiv:1308.0814 (2013).

\bibitem{She}
C. Shen,
Algebraic methods in sum-product phenomena,
{\it Israel J. Math.} 188(1) (2012), 123--130.

\bibitem{Soly}
J. Solymosi,
Bounding multiplicative energy by the sumset,
{\it Advances Math.} 222 (2009), 402--408.

\bibitem{SoT}
J. Solymosi and T. Tao,
An incidence theorem in higher dimension,
{\it Discrete Comput. Geom.} 48 (2012), 255--280. 

\bibitem{St}
Y. Stein,
The total reducibility order of a polynomial in two variables,
{\it Israel J. Math.} 68(1) (1989), 109--122.

\bibitem{Sz97}
L. Sz\'ekely,
Crossing numbers and hard Erd{\H o}s problems in discrete geometry,
{\it Combin. Probab. Comput.} 6 (1997), 353--358.

\bibitem{ST}
E. Szemer\'edi and W. Trotter,
Extremal problems in discrete geometry,
\emph{Combinatorica} 3(3) (1983), 381--392.

\bibitem{Tao}
T. Tao,
Expanding polynomials over finite fields of large characteristic, and a regularity lemma for definable sets,
in arXiv:1211.2894 (2011).

\bibitem{Th65}
R. Thom, 
Sur l'homologie des vari\'et\'es algebriques r\'eelles,
in: S.S. Cairns (ed.), {\it Differential and Combinatorial Topology}, 
Princeton University Press,
Princeton, NJ 1965, 255--265.

\bibitem{Tot}
C. D. T\'oth,
The Szemer\'edi-Trotter theorem in the complex plane,
in arXiv:0305283 (2003).

\bibitem{Ung}
P. Ungar,
$2N$ non-collinear points determine at least $2N$ directions,
{\em J. Combin. Theory Ser. A}, 33 (1982), 343--347.


\bibitem{Zah}
J. Zahl,
A Szemer\'edi-Trotter type theorem in $\mathbb{R}^ 4$,
in arXiv:1203.4600.


\bibitem{Zi89}
R. Zippel,
An explicit separation of relativised random polynomial time and relativised deterministic polynomial time,
{\it Inform. Process. Lett.} 33(4) (1989), 207--212.

\end{thebibliography}
\end{document}